\documentclass[superscriptaddress,amsmath,amssymb,nofootinbib,eqsecnum,a4paper,secnumarabic,11pt]{revtex4}
\pdfoutput=1
\usepackage{ascmac,braket,bm,mathrsfs,amsthm,amsfonts}
\usepackage{hyperref}
\usepackage{graphicx}
\usepackage{titlesec}
\usepackage{ulem}
\newcommand*{\justifyheading}{\raggedright}
\titleformat{\chapter}[display]
  {\normalfont\huge\bfseries\justifyheading}{\chaptertitlename\ \thechapter}
  {20pt}{\Huge}
\titleformat{\section}
  {\normalfont\Large\bfseries\justifyheading}{\thesection}{1em}{}
\titleformat{\subsection}
  {\normalfont\large\bfseries\justifyheading}{\thesubsection}{1em}{}
\titleformat{\subsubsection}
  {\normalfont\bfseries\justifyheading}{\thesubsubsection}{1em}{}
\usepackage[english]{babel}
\usepackage[dvipsnames]{xcolor}
\newcommand{\del}{\partial}
\newcommand{\nn}{\nonumber}
\newcommand{\tr}{\mbox{tr}}
\newcommand{\VEV}[1]{\langle #1 \rangle}
\newcommand{\akdel}{\overleftrightarrow\partial}

\begin{document}

\begin{flushright}
{\tt IPMU21-0047}
\end{flushright}

\title{Gamma-ray line from electroweakly interacting 
\\
non-abelian spin-1 dark matter
}

\author{Tomohiro Abe}
\email{abe.tomohiro@rs.tus.ac.jp}
\affiliation{Department of Physics, Faculty of Science and Technology, Tokyo University of Science, 2641 Yamazaki, Noda, Chiba 278-8510, Japan
}

\author{Motoko Fujiwara}
\email{motoko@eken.phys.nagoya-u.ac.jp}
\affiliation{
  Department of Physics, Nagoya University, Furo-cho Chikusa-ku, Nagoya, 464-8602 Japan
}

\author{Junji Hisano}
\email{hisano@eken.phys.nagoya-u.ac.jp}
\affiliation{
  Kobayashi-Maskawa Institute for the Origin of Particles and the
  Universe, Nagoya University,
  Furo-cho Chikusa-ku, Nagoya, 464-8602 Japan
}
\affiliation{
  Department of Physics, Nagoya University, Furo-cho Chikusa-ku, Nagoya, 464-8602 Japan
}
\affiliation{
  Kavli IPMU (WPI), UTIAS, University of Tokyo, Kashiwa, 277-8584, Japan
}

\author{Kohei Matsushita}
\email{kohei@eken.phys.nagoya-u.ac.jp}
\affiliation{
  Department of Physics, Nagoya University, Furo-cho Chikusa-ku, Nagoya, 464-8602 Japan
}

\begin{abstract}
We study gamma-ray line signatures from electroweakly interacting non-abelian spin-1 dark matter (DM). 
In this model, $Z_2$-odd spin-1 particles including a DM candidate have the SU(2)$_L$ triplet-like features, and the Sommerfeld enhancement is relevant in the annihilation processes. 
We derive the annihilation cross sections contributing to the photon emission and compare with the SU(2)$_L$ triplet fermions, such as Wino DM in the supersymmetric Standard Model. 
The Sommerfeld enhancement factor is approximately the same in both systems, while our spin-1 DM predicts the larger annihilation cross sections into $\gamma  \gamma/ Z  \gamma$ modes than those of the Wino by $\frac{38}{9}$. 
This is because a spin-1 DM  pair forms not only $J=0$ but also $J=2$ partial wave states where $J$ denotes the total spin angular momentum. 
Our spin-1 DM also has a new annihilation mode into $Z_2$-even extra heavy vector and photon, $Z'  \gamma$.
For this mode, 
the photon energy depends on the masses of DM and the heavy vector, 
and thus we have a chance to probe the mass spectrum.
The latest gamma-ray line search in the Galactic Center region gives a strong constraint on our spin-$1$ DM.
We can probe the DM mass for $\lesssim  25.3~$TeV by the Cherenkov Telescope Array experiment 
even if we assume a conservative DM density profile.
\end{abstract}

\maketitle

\clearpage
\tableofcontents

\section{Introduction}

We have overwhelming evidence that suggests 
dark matter (DM) in our universe. 
Although the nature of DM remains unrevealed, 
we have one quantitative piece of information about DM, 
the DM energy density in the current universe. 
This energy density is determined to be $\Omega  h^2  =  0.120  \pm  0.001$ 
by the Planck collaboration assuming the $\Lambda$CDM cosmological model~\cite{1807.06209}.

One of the most promising DM candidates is the Weakly Interacting Massive Particles (WIMPs). 
If we assume the WIMPs in our expanding universe, we can explain the current DM energy density as the thermal relic abundance through the Freeze-out mechanism~\cite{Lee:1977ua}. 
In this scenario, the velocity-weighted annihilation cross section for DM, $\Braket{\sigma  v_{\rm  rel}}$, 
characterizes the prediction of DM abundance. 
To reach the observed DM energy density, we need the value of the canonical cross section, $\Braket{\sigma  v_{\rm  rel}}  \simeq  3  \times  10^{-26}$~cm$^3$/s. 
This value is obtained once we assume DM mass and couplings to be the typical values of those in the electroweak theory in the Standard Model (SM). 
This fact is our motivation to consider DM candidates with electroweak interactions.

Spin-$0$ and spin-$1/2$ DM with electroweak interactions are systematically studied in the contexts of the Minimal DM~\cite{hep-ph/0512090,0706.4071,0903.3381}.\footnote{The references for the other models are given in Ref.~\cite{2105.11574}.}
The Minimal Supersymmetric SM predicts concrete DM candidates stabilized by the symmetry called R-parity, 
which is studied in many contexts~\cite{hep-ph/9906527,hep-ph/0601041,1410.4549,1601.04718}. 
The model of spin-$1$ DM with electroweak interactions is also studied in the extra-dimensional models~\cite{0811.1598,1803.01274,1702.02949,1808.10464}.

To reveal the general features of spin-$1$ DM, 
we construct the renormalizable four-dimensional model of electroweakly interacting non-abelian spin-$1$ DM~\cite{2004.00884}. 
This model is the minimal setup that realizes the fundamental features of the five-dimensional model of the spin-$1$ DM such as the $Z_2$-parity and the degenerated mass spectrum. 
We impose the exchange symmetry between the gauge groups to realize the above features, 
which is inspired by the method of deconstructing dimensions~\cite{hep-th/0104005,hep-th/0105239}. 
Our spin-$1$ DM couples to the electroweak gauge bosons with the electroweak gauge couplings,
and the DM mass should be $\gtrsim  {\cal  O} (1)$~TeV to explain the correct DM energy density. 
In this region, the electroweak bosons form the approximately long-range force potential.
Consequently, the Sommerfeld enhancement is relevant in the DM annihilation processes as discussed in the other electroweakly interacting DM models~\cite{hep-ph/0212022,hep-ph/0307216, hep-ph/0412403, 0810.0713, 1603.01383, 2105.07650}.
Due to this enhancement, monochromatic spectral gamma-ray lines from the DM annihilation are the striking signals for our spin-1 DM.

In this paper, we study gamma-ray line signatures from electroweakly interacting non-abelian spin-1 DM. 
We make a comparison with the pure Wino DM, which is the spin-$\frac{1}{2}$ DM candidate of the SU(2)$_L$ triplet in supersymmetric models, and clarify the differences between these systems. 
We derive the constraint from the latest gamma-ray observation in the Galactic Center region. 
We also reveal the region where we can probe in the future gamma-ray observation.

The rest of the paper is organized as follows. 
In Sec.~\ref{sec:model}, we briefly review the electroweakly interacting non-abelian spin-1 DM. 
In Sec.~\ref{sec:two-body_action}, we show the two-body effective action for the spin-$1$ DM system. 
We also show the annihilation cross section contributing to the monochromatic gamma-ray signals. 
In Sec.~\ref{sec:line_gamma-ray}, we present our numerical results. 
We compare the predicted annihilation cross section with that of the Wino DM and figure out distinctive features of our spin-$1$ DM. 
We show the constraints from the latest search for a monochromatic spectral line and the prospect sensitivity in the future gamma-ray observation. 
Our conclusions are given in Sec.~\ref{sec:conclusions}. 
We show the derivation of the two-body effective action in Appendix \ref{sec:derivation_S_eff}.

\section{Model}
\label{sec:model}

We briefly introduce a model of electroweakly interacting non-abelian vector DM~\cite{2004.00884}. 
We extend the electroweak gauge symmetry in the SM into SU(2)$_0\times$SU(2)$_1\times$SU(2)$_2\times$U(1)$_Y$.  
The gauge bosons for SU(2)$_0$, SU(2)$_1$, SU(2)$_2$, and U(1)$_Y$ are denoted as 
$W^a_{0\mu}$, $W^a_{1\mu}$, $W^a_{2\mu}$, and $B_\mu$, respectively ($a  =  1, 2, 3$). 
The gauge couplings for each symmetry are denoted as $g_0$, $g_1$, $g_2$, and $g'$, respectively. 
We summarize the matter fields and Higgs fields in Table \ref{tab:matter}.
Each fermion field corresponds to the SM fermion with the same SU(3)$_c$ and U(1)$_Y$ charge.
We introduce bi-fundamental scalar fields, $\Phi_j$ $( j  =  1, 2 )$, expressed as the two-by-two matrices. 
We impose the real conditions for $\Phi_j$ to reduce the degrees of freedom. 
\begin{align}
  \Phi_j = -\epsilon \Phi_j^* \epsilon, \quad
  \text{where} \quad 
  \epsilon = \begin{pmatrix} 0 & 1 \\ -1 & 0 \end{pmatrix}, 
\end{align}
and each of $\Phi_j$ contains four real degrees of freedom. 
The gauge transformations of $\Phi_j$ and $H$ are shown below.
\begin{align}
  \Phi_1  &\mapsto  U_0  \Phi_1  U_1^\dagger, 
  &
  \Phi_2  &\mapsto  U_2  \Phi_2  U_1^\dagger,
  &
  H  &\mapsto  e^{i  \theta_Y}  U_1  H,
\end{align}
where $U_n$ denote the two-by-two gauge transformation matrices of SU(2)$_n$ $( n  =  0,  1,  2 )$ and 
$\theta_Y$ is the phase of the U(1)$_Y$. 
We also impose a discrete symmetry under the following transformation.
\begin{align}
  &\Phi_1 \mapsto \Phi_2,
  &
  &\Phi_2 \mapsto \Phi_1,
  &
  &W_{0\mu}^a \mapsto   W_{2\mu}^a,
  &
  &W_{2\mu}^a \mapsto   W_{0\mu}^a,
\end{align}
where all the other fields remain unchanged. 
This transformation is equivalent to the exchange of SU(2)$_0$ and SU(2)$_2$, 
and thus it also requires $g_0 = g_2$.
%
\begin{table}[tb]
  \centering
  \caption{The matter and Higgs fields and their gauge charges. 
                The generation indices for the matter fields are implicit.}
  \label{tab:matter}
  
  \begin{tabular}{cc|ccccc}\hline
      field    & spin           & SU(3)$_c$ & SU(2)$_0$ & SU(2)$_1$ & SU(2)$_2$ & U(1)$_Y$ \\ \hline
      $q_L$    & $\frac{1}{2}$  &   3   &   1       &    2      &     1     &   $\frac{1}{6}$ 
      \\
      $u_R$    & $\frac{1}{2}$  &   3   &   1       &    1      &     1     &   $\frac{2}{3}$ 
      \\
      $d_R$    & $\frac{1}{2}$  &   3   &   1       &    1      &     1     &  $-\frac{1}{3}$ 
      \\
      $\ell_L$ & $\frac{1}{2}$  &   1   &   1       &    2      &     1     &  $-\frac{1}{2}$ 
      \\
      $e_R$    & $\frac{1}{2}$  &   1   &   1       &    1      &     1     &  $-1$ 
      \\ 
      \hline
      $H$      & 0              &   1   &   1       &    2      &     1     &   $\frac{1}{2}$ 
      \\
      $\Phi_1$ & 0              &   1   &   2       &    2      &     1     &   0 
      \\
      $\Phi_2$ & 0              &   1   &   1       &    2      &     2     &   0 
      \\ 
      \hline
  \end{tabular}
\end{table}

The Lagrangian for the extended bosonic sector is
\begin{align}
     {\cal L}\supset&
     - \frac{1}{4} B_{\mu \nu} B^{\mu \nu} 
     -\sum_{j=0}^2\sum_{a=1}^3 \frac{1}{4} W_{j\mu \nu}^a W_{j}^{a \mu \nu} 
     \nonumber\\
    & + D_{\mu}H^\dagger D^{\mu} H 
      + \frac{1}{2} \tr{D_{\mu} \Phi_1^\dagger D_{\mu} \Phi_1}
      + \frac{1}{2} \tr{D_{\mu} \Phi_2^\dagger D_{\mu} \Phi_2}
     \nonumber
     \\
     &
     - V_{\text{scalar}},
     \label{eq:Lagrangian}
\end{align}
where $V_{\text{scalar}}$ is the scalar potential as shown below. 
\begin{align}
     V_{\text{scalar}}
    =&
     m^2 H^\dagger H 
    + m_\Phi^2 
        \left[ 
          \tr\left(\Phi_1^\dagger \Phi_1\right)
          +   \tr\left(\Phi_2^\dagger \Phi_2\right)
        \right]
    \nonumber\\
    &
    + \lambda (H^\dagger H)^2
    + \lambda_\Phi 
        \left[
          \left(\tr\left(\Phi_1^\dagger \Phi_1\right)\right)^2
          +  \left(\tr\left(\Phi_2^\dagger \Phi_2\right)\right)^2
        \right]
    \nonumber\\
    &
    + \lambda_{h\Phi} H^\dagger H  
            \left[  
              \tr\left(\Phi_1^\dagger \Phi_1\right)
              +  \tr\left(\Phi_2^\dagger \Phi_2\right)
            \right]
    + \lambda_{12} \tr\left(\Phi_1^\dagger \Phi_1\right) \tr\left(\Phi_2^\dagger \Phi_2\right).
\end{align}
We assume the following vacuum expectation values (VEVs) to realize the U(1)$_{em}$ symmetry in the SM.
\begin{align}
 \VEV{\Phi_1}&= \VEV{\Phi_2}= \frac{1}{\sqrt{2}}\begin{pmatrix} v_\Phi & 0 \\ 0 & v_\Phi \end{pmatrix}, 
 &
  \VEV{H}=& \begin{pmatrix} 0 \\ \frac{v}{\sqrt{2}} \end{pmatrix}, 
\end{align}
where we take $v_\Phi  \gg  v$.
We define component fields around the VEVs. 
\begin{align}
 \Phi_j
 &= 
   \begin{pmatrix} 
    \frac{v_\Phi + \sigma_j + i \pi_j^0}{\sqrt{2}} &  i \pi_j^+ \\
    i \pi_j^-   &  \frac{v_\Phi + \sigma_j - i \pi_j^0}{\sqrt{2}}   
    \end{pmatrix}, 
 &
 H=& \begin{pmatrix} i\pi_3^+ \\ \frac{v + \sigma_3 - i \pi_3^0}{\sqrt{2}} \end{pmatrix}.
\end{align}
After $\Phi_1$ and $\Phi_2$ develop the nonzero VEVs, we still have the residual SU(2)$\times$U(1)$_Y$  gauge symmetry whose gauge transformation is shown below. 
\begin{align}
  \Phi_1  &\mapsto  U  \Phi_1  U^\dagger, 
  &
  \Phi_2  &\mapsto  U  \Phi_2  U^\dagger,
  &
  H  &\mapsto  e^{i  \theta_Y }  U  H, 
\end{align}
where $U$ is the SU(2) two-by-two gauge transformation matrix. 
This SU(2) transformation corresponds to the SU(2)$_0\times$SU(2)$_1\times$SU(2)$_2$ transformation with $U_0  =  U_1  =  U_2  \equiv  U$. 
This SU(2) is regarded as the SU(2)$_L$ in the SM and simply called the SU(2)$_L$ in the following discussion.
The SU(2)$_L$$\times$U(1)$_Y$ symmetry is broken by $\VEV{H}$ 
to the U(1)$_{em}$ symmetry.

After all the Higgs fields develop nonzero VEVs, we still have the exchange symmetry whose symmetry transformations are shown below. 
\begin{align}
  &\sigma_1 \mapsto \sigma_2,
  &
  &\sigma_2 \mapsto \sigma_1,
  &
  &W_{0\mu}^a \mapsto   W_{2\mu}^a,
  &
  &W_{2\mu}^a \mapsto   W_{0\mu}^a.
  \label{eq:exchange_SSB}
\end{align}
To find out a $Z_2$ parity from this residual discrete symmetry, we anti-symmetrize the exchanged fields and define the following states. 
\begin{align}
  h_D  &\equiv  \frac{\sigma_1  -  \sigma_2}{\sqrt{2}},
  \\
  V^0_\mu  &\equiv  \frac{W^0_{0\mu}  -  W^0_{2\mu}}{\sqrt{2}},
  \\
  V^\pm_\mu  &\equiv  \frac{W^\pm_{0\mu}  -  W^\pm_{2\mu}}{\sqrt{2}},
\end{align}
where
\begin{align}
  W^\pm_{n\mu}  =  \frac{W^1_{n\mu}  \mp  i  W^2_{n\mu}}{\sqrt{2}}
  ~~~~
  ( n  =  0, 2 ). 
\end{align}
These fields are eigenstates of the U(1)$_{em}$ charge.\footnote{The U(1)$_{em}$ symmetry generator is expressed as $Q  =  T^3_0  +  T^3_1  +  T^3_2  +  Y$ where $T^3_n$ denote the third generators of SU(2)$_n$ ($n  =  0,  1,  2$).}
These states acquire $(-1)$ factors under the transformation in Eq.~\eqref{eq:exchange_SSB} while all the other states remain unchanged. 
This is nothing but a $Z_2$ parity assignment. 
Note that the $Z_2$-odd states are automatically mass eigenstates since the mass mixing terms with other states are forbidden by this $Z_2$ parity.

We refer to $Z_2$-odd spin-1 particles, $V^0$ and $V^\pm$, as ``\textit{$V$-particles}."
The $V$-particles are approximately regarded as the spin-$1$ triplet of SU(2)$_L$ and have the degenerated masses at the tree-level. 
\begin{align}
  m_{V^0}^2  =  m_{V_\pm}^2  =  \frac{g_0^2  v_\Phi^2}{4}  \equiv  m_V^2. 
\end{align}
The electroweak radiative corrections break the degeneracy, which makes the charged component slightly heavier than the neutral one. 
We find the following value for mass splitting at the one-loop level.
\begin{align}
  \delta  m_V  \equiv  m_{V_\pm}  -  m_{V_0}  \simeq  168~\text{MeV}. 
\end{align}
We assume $h_D$ is heavier than $V^0$ to focus on the spin-$1$ DM scenario.  
Therefore, $V^0$ is the lightest $Z_2$-odd particle and our stable spin-1 DM candidate.

\subsection{Physical spectrum}
\label{sec:physical_spectrum}

We summarize the physical spectra in our model. 
Diagonalizing the mass matrices, we obtain the following mass eigenstates in the Higgs sector.
\begin{align}
  \begin{pmatrix}
    h_D  \\  h \\ h'
  \end{pmatrix}
  =
  \begin{pmatrix}
    1  &  0  &  0\\
    0  &  \sin\phi_h  &  \cos\phi_h\\
    0  &  \cos\phi_h  &  -  \sin\phi_h
  \end{pmatrix}
  \begin{pmatrix}
    \frac{1}{\sqrt{2}}  &  -  \frac{1}{\sqrt{2}}  &  0\\
    \frac{1}{\sqrt{2}}  &  \frac{1}{\sqrt{2}}  &  0\\
    0  &  0  &  1 \\
  \end{pmatrix}
  \begin{pmatrix}
    \sigma_1 \\ \sigma_2  \\  \sigma_3
  \end{pmatrix}.
\end{align}
We take the scalar masses, $\{ m_h, m_{h'}, m_{h_D} \}$, and the mixing angle, $\phi_h$, as input parameters. 
We can express the dimensionless couplings of $V_{\rm  scalar}$ in these input parameters as shown below.
\begin{align}
  \lambda=& \frac{m_h^2 \cos^2\phi_h + m_{h'}^2 \sin^2\phi_h}{2 v^2}, \label{eq:lambda}\\
  \lambda_{h\Phi}=& -\frac{\sin\phi_h \cos\phi_h}{2 \sqrt{2} v v_\Phi} (m_{h'}^2 - m_h^2),\\
  \lambda_{\Phi}=& \frac{m_h^2 \sin^2\phi_h + m_{h'}^2 \cos^2\phi_h + m_{h_D}^2}{16 v_\Phi^2},\\
  \lambda_{12}=& \frac{m_h^2 \sin^2\phi_h + m_{h'}^2 \cos^2\phi_h - m_{h_D}^2}{8 v_\Phi^2}. 
  \label{eq:lambda12}
\end{align}
In the following discussion, we focus on $\phi_h  \lesssim  0.1$ to evade the direct detection constraints~\cite{2004.00884}.

For the charged gauge bosons, we obtain  
\begin{align}
 \begin{pmatrix}
  V^{\pm}_\mu \\  W^{\pm}_\mu \\ W'^{\pm}_{\mu}
 \end{pmatrix}
&=
  \begin{pmatrix}
    1  &  0  &  0 \\
    0  &  \cos\phi_\pm  &  \sin\phi_\pm \\
    0  &  -  \sin\phi_\pm  &  \cos\phi_\pm
  \end{pmatrix}
  \begin{pmatrix}
    \frac{1}{\sqrt{2}}  &  0  &  -  \frac{1}{\sqrt{2}} \\
    0  &  1  &  0 \\
    \frac{1}{\sqrt{2}}  &  0    &  \frac{1}{\sqrt{2}}
  \end{pmatrix}
 \begin{pmatrix}
  W_{0\mu}^{\pm} \\ W_{1\mu}^{\pm} \\ W_{2\mu}^{\mu}
 \end{pmatrix},
\end{align}
where 
\begin{align}
  \cos\phi_\pm = \sqrt{\frac{m_{V^\pm}^2 - m_W^2}{m_{W'}^2 - m_W^2}},
  \quad
  \sin\phi_\pm = \sqrt{\frac{m_{W'}^2 - m_{V^\pm}^2}{m_{W'}^2 - m_W^2}}.
\end{align}
The masses of charged $Z_2$-even vectors are obtained as 
\begin{align}
  m_W^2
  =&
  \frac{1}{8}
  \left\{
    g_1^2 v^2 + (g_0^2 + 2 g_1^2) v_\Phi^2
    -  \sqrt{-4 g_0^2 g_1^2 v^2 v_\Phi^2  +  \left[ g_1^2 v^2 + (g_0^2 + 2 g_1^2 ) v_\Phi^2 \right]^2}
  \right\},
  \\
  m_{W'}^2
  =&
  \frac{1}{8}
  \left\{
    g_1^2 v^2 + (g_0^2 + 2 g_1^2) v_\Phi^2  
    +  \sqrt{-4 g_0^2 g_1^2 v^2 v_\Phi^2  +  \left[ g_1^2 v^2 + (g_0^2 + 2 g_1^2 ) v_\Phi^2 \right]^2}
  \right\}.
\end{align}

For the neutral gauge bosons, we obtain
\begin{align}
  \begin{pmatrix}
    V^{0}_\mu \\  A_\mu \\ Z_\mu \\ Z'_{\mu}
  \end{pmatrix}
  &=
  \begin{pmatrix}
    \frac{1}{\sqrt{2}} & 0 & - \frac{1}{\sqrt{2}} & 0 
    \\
    \frac{e}{g_0} &  \frac{e}{g_1} &  \frac{e}{g_0} &  \frac{e}{g'} 
    \\
    \omega^0_{Z} & \omega^1_Z & \omega^0_Z & \omega^B_Z
    \\
    \omega^0_{Z'} & \omega^1_{Z'} & \omega^0_{Z'} & \omega^B_{Z'}
    \\
  \end{pmatrix}
  \begin{pmatrix}
    W_{0\mu}^{3} \\ W_{1\mu}^{3} \\ W_{2\mu}^{3} \\ B_\mu
  \end{pmatrix},
\end{align}
where
\begin{align}
    e =& \left( \frac{2}{g_0^2} + \frac{1}{g_1^2} + \frac{1}{g'^2} \right)^{-1/2},
    \\
    \omega^0_Z
    =& 
    \frac{e g_1}{\sqrt{g_0^2 + 2 g_1^2} g'} \cos\phi_0
    +
    \frac{g_0}{\sqrt{2(g_0^2 + 2 g_1^2)}} \sin\phi_0
    ,\\
    \omega^1_Z
    =& 
    \frac{e g_0}{\sqrt{g_0^2 + 2 g_1^2} g'} \cos\phi_0
    -
    \frac{\sqrt{2} g_1}{\sqrt{g_0^2 + 2 g_1^2}} \sin\phi_0
    ,\\
    \omega^B_Z
    =& 
    -\frac{e \sqrt{g_0^2 + 2 g_1^2}}{g_0 g_1} \cos\phi_0
    ,\\
    \omega^0_{Z'}
    =& 
    \frac{g_0}{\sqrt{2(g_0^2 + 2 g_1^2)}} \cos\phi_0
    -
    \frac{e g_1}{\sqrt{g_0^2 + 2 g_1^2} g'} \sin\phi_0
    ,\\
    \omega^1_{Z'}
    =& 
    -\frac{\sqrt{2} g_1}{\sqrt{g_0^2 + 2 g_1^2}} \cos\phi_0
    -\frac{e g_0}{\sqrt{g_0^2 + 2 g_1^2} g'} \sin\phi_0
    ,\\
    \omega^B_{Z'}
    =& 
    \frac{e \sqrt{g_0^2 + 2 g_1^2}}{g_0 g_1} \sin\phi_0
    .
\end{align}
We define $\phi_0$ that satisfies the following relation. 
\begin{align}
  \frac{1}{4}
  \begin{pmatrix}
    \frac{g_0^2 g_1^2 g'^2}{e^2 (g_0^2 + 2 g_1^2)} v^2   
    &  -\frac{\sqrt{2} g_0 g_1^3 g'}{e (g_0^2 + 2 g_1^2)} v^2 
    \\  
    -\frac{\sqrt{2} g_0 g_1^3 g'}{e (g_0^2 + 2 g_1^2)} v^2 
    &  \frac{(g_0^2 + 2 g_1^2)^2 v_\Phi^2 + 2 g_1^4 v^2}{(g_0^2 + 2 g_1^2)}
  \end{pmatrix}
  \begin{pmatrix}
    \cos\phi_0 &  -\sin\phi_0 
    \\ 
    \sin\phi_0 &  \cos\phi_0 
  \end{pmatrix}
  =&
  \begin{pmatrix}
    \cos\phi_0 &  -\sin\phi_0 
    \\ 
    \sin\phi_0 &  \cos\phi_0 
  \end{pmatrix}
  \begin{pmatrix}
    m_Z^2 & 0 
    \\ 
    0 & m_{Z'}^2
  \end{pmatrix}.
\end{align}

The parameters in the Lagrangian in Eq.~\eqref{eq:Lagrangian} are summarized below. 
\begin{align}
    \left\{
      g_0,\ g_1,\ g',\ m^2,\ m_\Phi^2,\ 
      \lambda,\ \lambda_\Phi,\ \lambda_{h\Phi}, \lambda_{12}
    \right\}.
\label{eq:parameters}
\end{align}
We choose the following input parameters for our phenomenological study.
\begin{align}
  \left\{
    e,\ m_Z,\ v,\ m_h,\ m_{Z'},\ m_V,\ 
     m_{h'},\ m_{h_D}, \phi_h
  \right\}.
\label{eq:inputs}
\end{align}
The first four parameters are already fixed by the experiments, while the last five parameters are free.
As we mentioned before, 
we consider the small $\phi_h$ regime. 
Therefore, 
the SM Higgs coupling with the $V$-particles are suppressed by the small mixing factor, $\sin  \phi_h$. 
Although we do not specify the values of $m_{h'}$ and $m_{h_D}$ in the following discussion, 
we assume these scalar masses are in the TeV scale to focus on the phenomenological aspects resulting from the electroweak interactions. 
The constraints on these parameters in the scalar sector 
are studied in Ref.~\cite{2004.00884}.
For later convenience, we introduce $g_W$ as defined below.
\begin{align}
 g_W
 \equiv
 \left( \frac{2}{g_0^2} + \frac{1}{g_1^2}\right)^{-1/2}.
\end{align}
This coupling corresponds to the SU(2)$_L$ gauge coupling approximately, and we obtain $g_W \simeq 0.65$, which is the same value as the SU(2)$_L$ gauge coupling in the SM, in the limit of $v_\Phi  \gg  v$.

We assume $v_\Phi  \gg  v$ throughout our study, and it is useful to derive the approximated forms for the physical values in this limit. 
For $m_{W}$ and $m_{W'}$, we obtain
\begin{align}
  m_{W}  &\simeq  \frac{g_W  v}{2}, 
  \\
  m_{W'}  &\simeq  m_{Z'}, 
\end{align}
and thus we easily obtain the correct value of $m_W$. 
The gauge couplings, $g_0$ and $g_1$, are expressed as 
\begin{align}
  g_0  \simeq&  \sqrt{2} g_W \frac{m_{Z'}}{m_V} \frac{1}{\sqrt{\frac{m_{Z'}^2}{m_V^2}-1}}, 
  \label{eq:g0}
  \\
  g_1  \simeq&  g_W \frac{m_{Z'}}{m_V}. 
  \label{eq:g1}
\end{align}
These couplings are constrained by the perturbative unitarity bounds. 
We obtain the following constraints in the high energy regime~\cite{1202.5073}.
\begin{align}
  g_j  <  \sqrt{\frac{16  \pi}{\sqrt{6}}}  \simeq  4.53~~~~( j  =  0, 1 ).
\end{align}
From these bounds, we can narrow down the viable range of the mass ratio, $\frac{m_{Z'}}{m_V}$. 
If we take $g_W  \simeq  0.65$, we obtain 
\begin{align}
  1.02  \lesssim  \frac{m_{Z'}}{m_V}  \lesssim  6.97. 
  \label{eq:unitarity_g0}
\end{align}
More detailed explanations of our model are given in Ref.~\cite{2004.00884}.

\subsection{Couplings of $V$-particles}
\label{sec:couplings}

We show the couplings of the $V$-particles in the limit of $v_\Phi  \gg  v$. 
The vector triple couplings are shown below.
\begin{align}
  {\cal  L}  
  &\supset
  i  C_{XYZ}
  \biggl[
    \left( X_\nu^+ \akdel^\mu  Y^{-  \nu} \right)  Z_\mu^0  
    +
    \left( Y_\nu^- \akdel^\mu  Z^{0  \nu} \right)  X_\mu^+
    +  
    \left( Z_\nu^0 \akdel^\mu  X^{+  \nu} \right)  Y_\mu^-
  \biggr], 
\end{align} 
where
\begin{align}
  C_{W^+V^-V^0}  =  C_{V^+W^-V^0}  &\simeq  g_W, 
  \\
  C_{V^+  V^-  A}  &=  e,
  \\
  C_{V+  V-  Z}  &\simeq  g_W  c_W,
  \\
  C_{V^+  V^-  Z'}  &\simeq  \frac{g_W}{\sqrt{\frac{m_{Z'}^2}{m_V^2}  -  1}}  \equiv  g_{Z'}.
  \label{eq:g_Zp}
\end{align}
We define $s_W  \equiv  \sin  \theta_W$ and $c_W  \equiv  \cos  \theta_W$ where $\theta_W$ is the Weinberg angle.
We also define $g_{Z'}$ in Eq.~\eqref{eq:g_Zp} to characterize the couplings between $Z'$ and the  $V$-particles.
The vector quartic couplings are shown below. 
\begin{align}
  {\cal  L}  
  &\supset
  C_{XYZW}  ~
  X^+_\rho  Y^-_\sigma  Z^0_\mu  W^0_\nu
  \left(  
        g^{\rho  \mu}  g^{\sigma  \nu}    
    +  g^{\rho  \nu}  g^{\sigma  \mu}
    -  2  g^{\rho  \sigma}  g^{\mu  \nu}
  \right),
\end{align}
where 
\begin{align}
  C_{V^+  V^-  A  A}  &=  \frac{e^2}{2},
  \\
  C_{V^+  V^-  A  Z}  &\simeq  e  g_W  c_W,
  \\
  C_{V^+  V^-  A  Z'}  &=  e  C_{V^+  V^-  Z'}.
\end{align}

\section{Annihilation cross section for $V$-particles}
\label{sec:two-body_action}

As studied in the previous work~\cite{2004.00884}, we need $m_{V}  \gtrsim  {\cal  O}  (1)$~TeV to explain the correct DM relic abundance in our model. 
In this region, the $W$ and $Z$ bosons effectively form the long-range force potential for the $V$-particles. 
Consequently, we have the sizable Sommerfeld enhancement in the DM annihilation processes. 
For this class of models, the monochromatic gamma-ray line from the DM annihilation is an excellent probe.

In this section, we summarize the formulas to discuss the gamma-ray line signature from our spin-$1$ DM. 
We apply the same formalism as those for the spin-$0$ or spin-$1/2$ DM with electroweak interactions~\cite{hep-ph/0412403,0706.4071}. 
First, we define the non-relativistic (NR) field operators for the $V$-particles to fix our notations. 
Second, we show the NR two-body effective action for the $V$-particles. 
We focus on the electrically neutral two-body states to study the DM annihilation signals. 
We derive the effective action in Appendix~\ref{sec:derivation_S_eff}.

\subsection{Non-relativistic field operators for $V$-particles}

The asymptotic field operators for the $V$-particles are defined below. 
\begin{align}
  V^0_\mu  (x)
  &=  \sum_{A  =  1}^3  \int  \frac{d^3  p}{( 2  \pi )^3}  \frac{1}{\sqrt{2  E_{\rm{p}}}}  
         \left[ 
               a^A  ( \bm{p} )  e^{-  i  p  \cdot  x}  \epsilon^A_\mu  ( p )
           +  a^{A  \dagger}  ( \bm{p} )  e^{i  p  \cdot  x}  \epsilon^{A  *}_\mu  ( p )
         \right],
  \\
  V^-_\mu  (x)
  &=  \sum_{A  =  1}^3  \int  \frac{d^3  p}{( 2  \pi )^3}  \frac{1}{\sqrt{2  E_{\rm{p}}}}  
         \left[ 
               b^A  ( \bm{p} )  e^{-  i  p  \cdot  x}  \epsilon^A_\mu  ( p )
           +  d^{A  \dagger}  ( \bm{p} )  e^{i  p  \cdot  x}  \epsilon^{A  *}_\mu  ( p )
         \right],
  \\
  V^+_\mu  (x)
  &=  \sum_{A  =  1}^3  \int  \frac{d^3  p}{( 2  \pi )^3}  \frac{1}{\sqrt{2  E_{\rm{p}}}}  
         \left[ 
               d^A  ( \bm{p} )  e^{-  i  p  \cdot  x}  \epsilon^A_\mu  ( p )
           +  b^{A  \dagger}  ( \bm{p} )  e^{i  p  \cdot  x}  \epsilon^{A  *}_\mu  ( p )
         \right], 
\end{align}
where $\epsilon_\mu^A  (p)$ denote the physical polarization vectors.
These polarization vectors satisfy the transverse conditions, 
\begin{align}
  p^\mu  \epsilon_\mu^A  ( p )  =  0,
\end{align}
and the orthogonal relations. 
\begin{align}
  \epsilon_\mu^{A  *}  ( p )  \epsilon^{B  \mu}  ( p )  =  -  \delta^{A  B}. 
\end{align}
We impose the canonical commutation relations between the annihilation and creation operators.
\begin{align}
  [  a^A  ( \bm{p} ),  a^{B  \dagger}  ( \bm{p}' )  ]  &=  ( 2  \pi )^3  \delta^3  ( \bm{p}  -  \bm{p}' )  \delta^{A  B},
  \\
  [  b^A  ( \bm{p} ),  b^{B  \dagger}  ( \bm{p}' )  ]  &=  ( 2  \pi )^3  \delta^3  ( \bm{p}  -  \bm{p}' )  \delta^{A  B},
  \\
  [  d^A  ( \bm{p} ),  d^{B  \dagger}  ( \bm{p}' )  ]  &=  ( 2  \pi )^3  \delta^3  ( \bm{p}  -  \bm{p}' )  \delta^{A  B}. 
\end{align}

To derive the NR two-body effective action, we perform the NR expansion for these operators and integrate out the large momentum modes. 
Since the zeroth component of the polarization tensor is sub-leading, ${\cal  O}  ( |\bm{p}|/m_V )$, we focus on the spacial components ($i  =  1,  2,  3$) for the $V$-particle operators. 
\begin{align}
  V^0_i  (x)
  &\simeq
  \frac{1}{\sqrt{2  m_V}}
  \left[
    e^{-  i  m_V  t}
    {\cal  A}_i  (x)
    +
    e^{i  m_V  t}    
    {\cal  A}_i^\dagger  (x)
  \right],
  \label{eq:NR_V_def-0}
  \\
  V^-_i  (x)
  &\simeq
  \frac{1}{\sqrt{2  m_V}}
  \left[
    e^{-  i  m_V  t}
    {\cal  B}_i  (x)
    +
    e^{i  m_V  t}    
    {\cal  D}_i^\dagger  (x)
  \right],
  \\
  V^+_i  (x)
  &\simeq
  \frac{1}{\sqrt{2  m_V}}
  \left[
    e^{-  i  m_V  t}
    {\cal  D}_i  (x)
    +
    e^{i  m_V  t}    
    {\cal  B}_i^\dagger  (x)
  \right],
  \label{eq:NR_V_def-+}
\end{align}
where we define 
\begin{align}
  {\cal  A}_i  (x)
  &=
   \sum_{A  =  1}^3  \int  \frac{d^3  p}{( 2  \pi )^3}    
               a^A  ( \bm{p} )  e^{-  i  p  \cdot  x}  \epsilon^A_i  (p),
  \label{eq:NR_operator-A}
  \\
  {\cal  B}_i  (x)
  &=  \sum_{A  =  1}^3  \int  \frac{d^3  p}{( 2  \pi )^3}
               b^A  ( \bm{p} )  e^{-  i  p  \cdot  x}  \epsilon^A_i  (p),
  \label{eq:NR_operator-B}
  \\
  {\cal  D}_i  (x)
  &=  \sum_{A  =  1}^3  \int  \frac{d^3  p}{( 2  \pi )^3}
               d^A  ( \bm{p} )  e^{-  i  p  \cdot  x}  \epsilon^A_i  (p).
  \label{eq:NR_operator-D}
\end{align}
We also define the NR one particle states for the $V$-particles with the momentum $\bm{p}$ and the polarization $A$.
\begin{align}
  \Ket{ V^0; \bm{p}, A }  &\simeq  \sqrt{2  m_V}  a^{A  \dagger}  ( \bm{p})  \Ket{0}, 
  \\
  \Ket{ V^-; \bm{p}, A }  &\simeq  \sqrt{2  m_V}  b^{A  \dagger}  ( \bm{p})  \Ket{0}, 
  \\
  \Ket{ V^+; \bm{p}, A }  &\simeq  \sqrt{2  m_V}  d^{A  \dagger}  ( \bm{p})  \Ket{0}. 
\end{align}

\subsection{Two-body effective action}
\label{sec:two-body}

We show the two-body effective action for the NR $V$-particles. 
The electrically neutral two-body states are composed of $V^-  ( t,  \bm{x} )  V^+  ( t,  \bm{y} )$ or $V^0  ( t,  \bm{x} )  V^0  ( t,  \bm{y} )$.
We change the space-time coordinates into the center of mass coordinate, $R  =  ( R^0,  \bm{R} )$, and the relative coordinate, $\bm{r}$.
\begin{align}
  R^0  &=  t,
  &
  \bm{R}  &\equiv  \frac{\bm{x}  +  \bm{y}}{2},
  &
  \bm{r}  &\equiv  \bm{x}  -  \bm{y}. 
\end{align}
As shown later, the NR leading-order potential has a spherically symmetric form. 
Therefore, the total spin angular momentum, $J$, and $z$-component of the spin angular momentum, $J_z$, are good quantum numbers.
We obtain the following effective action in the decomposed form into each partial wave state.
\begin{align}
  S_{\rm  eff}
  &=  
  \sum_{J,  J_z}
  \int  d^4  R  d^3  r 
  ~ 
  \Phi^{J, J_z \dagger}  ( R,  \bm{r} )  \cdot  
  \left[ 
    \left( i  \del_{R^0}  +  \frac{\nabla_R^2}{4  m_V}  +  \frac{\nabla_r^2}{m_V} \right)
    -
    \hat{V}  (r)
    +  i  \frac{9}{2}  \hat{\Gamma}^{J}  \delta^3( \bm{r} )
  \right] 
  \cdot  \Phi^{J, J_z}  ( R,  \bm{r} ),
  \label{eq:SII}
\end{align}
where $r  \equiv  | \bm{r} |$ and $J  =  0,  1,  2$ with $|J_z|  \leq  J$.
The $\hat{V}$ and $\hat{\Gamma}^J$ denote the real part and imaginary part of the potential, respectively.
We define the two-body fields for each $(J, J_z)$ as shown below. 
\begin{align}
  \Phi^{J, J_z}  ( R,  \bm{r} )
  &=  
  \begin{pmatrix}
    \phi_C^{J, J_z}  ( R,  \bm{r} )
    \\
    \phi_N^{J, J_z}  ( R,  \bm{r} )
  \end{pmatrix}.
\end{align}
The upper and lower components are formed by $V^-  V^+$ and $V^0  V^0$, respectively.
\begin{align}
  \phi_C^{J,  J_z}  ( R,  \bm{r} )
  &=  
  {\cal  B}_i  ( R^0,  \bm{R}  +  \bm{r}/2 )~
  S_{ij}^{J,  J_z}~
  {\cal  D}_j  ( R^0,  \bm{R}  -  \bm{r}/2 )
  &&
  (J = 0, 1, 2),
  &
  \label{eq:phi_C}
  \\
  \phi_N^{J,  J_z}  ( R,  \bm{r} )
  &=  
  \frac{1}{\sqrt{2}}
  {\cal  A}_i  ( R^0,  \bm{R}  +  \bm{r}/2 )~
  S_{ij}^{J,  J_z}~
  {\cal  A}_j  ( R^0,  \bm{R}  -  \bm{r}/2 )
  &&
  (J = 0, 2).
  &
  \label{eq:phi_N}
\end{align}
In the above definition, we introduce the basis of three-by-three matrices. 
\begin{align}
  \hat{S}^{J,J_z} \equiv S^{J,J_z}_{ij} ~~~~(i,j = 1,2,3), 
\end{align}
where
\begin{align}
    \hat{S}^{0,0} &= \frac{-1}{\sqrt3} \begin{pmatrix} 1&0&0 \\ 0&1&0 \\ 0&0&1 \end{pmatrix}, 
    &&
    &&
    \label{eq:Sij_def-0}
    \\
    \nn
    \\
    \hat{S}^{1,1} &= \frac12 \begin{pmatrix} 0&0&-1 \\ 0&0&-i \\ 1&i&0 \end{pmatrix}, 
    &
    \hat{S}^{1,0} &= \frac1{\sqrt2} \begin{pmatrix} 0&i&0 \\ -i&0&0 \\ 0&0&0 \end{pmatrix}, 
    &
    \hat{S}^{1,-1} &= \frac12 \begin{pmatrix} 0&0&-1 \\ 0&0&i \\ 1&-i&0 \end{pmatrix},
    \label{eq:Sij_def-1}
    \\
    \nn
    \\
    \hat{S}^{2,2} &= \frac12 \begin{pmatrix} 1&i&0 \\ i&-1&0 \\ 0&0&0 \end{pmatrix},
    &
    \hat{S}^{2,1} &= \frac12 \begin{pmatrix} 0&0&-1 \\ 0&0&-i \\ -1&-i&0 \end{pmatrix},
    &
    \hat{S}^{2,0} &= \frac1{\sqrt6} \begin{pmatrix} -1&0&0 \\ 0&-1&0 \\ 0&0&2 \end{pmatrix},
    \nn
    \\
    \hat{S}^{2,-1} &= \frac12 \begin{pmatrix} 0&0&1 \\ 0&0&-i \\ 1&-i&0 \end{pmatrix},
    &
    \hat{S}^{2,-2} &= \frac12 \begin{pmatrix} 1&-i&0 \\ -i&-1&0 \\ 0&0&0 \end{pmatrix}.
    \label{eq:Sij_def-2}
\end{align}
These matrices satisfy the following orthogonal relation. 
\begin{align}
  \tr{\left[ \hat{S}^{J,J_z}  \hat{S}^{J',J'_z  *} \right]}  =  (-1)^J  \delta^{J  J'}  \delta^{J_z  J'_z}. 
\end{align}
Note that $\hat{S}^{J,J_z}$ is symmetric  matrices for $J=0,2$ and anti-symmetric for $J=1$. 
We have no $\phi^{1, J_z}_N$ because $\phi_N$ is composed of identical particles.
As mentioned before, $J$ is conserved as long as we focus on the NR leading order contributions. 
Therefore, $J=1$ partial wave modes are irrelevant to discuss the DM annihilation signals. 
We only consider $J=0, 2$ states in the following discussion. 
The normalization factors for the two-body states are fixed to realize the canonical weights for the two-body propagators. 
The two-body propagator of $\phi_C^{J, J_z}$ is defined as the time-ordered product.
\begin{align}
  \Braket{ 0  | T  \phi_C^{J, J_z}  ( R,  \bm{r} )  \phi_C^{J,  J_z  \dagger}  ( 0,  \bm{0} )  | 0 }
  &=
  \int  \frac{d^3  P}{( 2  \pi )^3}  
  \frac{d^3  k}{( 2  \pi )^3}  
  ~e^{ -  i  \Bigl( \frac{|\bm{P}|^2}{4  m}  +  \frac{\bm{k}^2}{m} \Bigr)  R^0  +  i  \bm{P}  \cdot  \bm{R} }  
  ~e^{ +  i  \bm{k}  \cdot  \bm{r} }  
  ~\theta  (R^0),
  \label{eq:2-body_prop}
\end{align}
where $\theta(R^0)$ is the Heaviside step function.
This expression is obtained by substituting the explicit forms of $\phi_C^{J, J_z}$ and $\phi_C^{J,  J_z  \dagger}$ and using the canonical commutation relation between them.\footnote{Note that $(\hat{S}^{J,  J_z})^\dagger  =  (-1)^J  (\hat{S}^{J,  J_z})^*$ because it is symmetric for $J  =  0,  2$ but anti-symmetric for $J  =  1$.}
The definition of $\phi^{J, J_z}_N$ has another normalization factor of $\frac{1}{\sqrt{2}}$ because $\phi^{J, J_z}_N$ is composed of identical particles.

The leading-order expression for $\hat{V}$ is obtained as follows.
\begin{align}
  \hat{V}  (r)  
  &=  
  \begin{pmatrix}
    \displaystyle
    2  \delta  m_V  -  \frac{\alpha_2  s_W^2}{r}  -  \frac{\alpha_2  c_W^2  e^{-  m_Z  r}}{r}  
    &  
    ~~~~
    \displaystyle    
    -  \frac{\sqrt{2}  \alpha_2  e^{-  m_W  r}}{r}
    \\
    \\
    \displaystyle    
    -  \frac{\sqrt{2}  \alpha_2  e^{-  m_W  r}}{r}  
    &
    ~~~~
    \displaystyle        
    0
  \end{pmatrix}
  ,
  \label{eq:V_0-2} 
\end{align}
where $\alpha_2  \equiv  \frac{g_W^2}{4  \pi}$. 
This potential is induced by the NR processes between $V$-particles. 
The off-diagonal elements are induced by the $W$ boson exchange processes, which cause the mixing between $\phi_C$ and $\phi_N$.
Compared with the leading-order contributions, the Higgs exchange contributions are suppressed by small $\phi_h$. 
The $W'$ and $Z'$ exchange contributions are exponentially suppressed by $m_{W'}$ and $m_{Z'}$. 
The contributions from the vector quadratic couplings are suppressed by $\frac{1}{m_V^2}$. 
The leading-order expression for $\hat{V}$ is the same as that of the SU(2)$_L$ triplet DM with spin-$0$/spin-$\frac{1}{2}$.  
This is because the spin-dependent features decouple from the NR processes.
We show the derivation of $\hat{V}$ in Appendix~\ref{sec:potential_Re}.

The imaginary part of the potential, $\hat{\Gamma}^J$, is derived by operator matching between the two-body field operators and the calculations of the loop amplitudes of the $V$-particles.\footnote{We choose the factor of ``$\frac{9}{2}$" in Eq.~\eqref{eq:SII} so that the (1,1)-component of $\hat{\Gamma}^J$ is equal to the tree-level spin-averaged velocity-weighted annihilation cross sections of $V^-  V^+$ with $J$.}
We focus on the one-loop diagrams with the intermediate states, $XX'  =  \gamma  \gamma,  Z  \gamma,  Z'  \gamma$, involving the gamma-ray line signatures. 
The annihilation modes into $h  \gamma$ and $h' \gamma$ only induce $J=1$ partial wave contributions, 
which is irrelevant to the DM annihilation signals.
We decompose $\hat{\Gamma}^{J}$ into each contribution of the intermediate state, $X  X'$, as denoted by $\Gamma^{J}_{XX'}$.
The leading-order expressions are shown below. 
\begin{align}
  \hat{\Gamma}^{J=0}_{\gamma  \gamma}
  &=
  \frac{2}{3}  \frac{\pi  \alpha_2^2}{m_V^2}
  \begin{pmatrix}
    s_W^4  &  0
    \\
    0  &  0
  \end{pmatrix},
  \label{eq:Gamma_GG-0}
  \\
  \hat{\Gamma}^{J=2}_{\gamma  \gamma}
  &=
  \frac{32}{45}  \frac{\pi  \alpha_2^2}{m_V^2}
  \begin{pmatrix}
    s_W^4  &  0
    \\
    0  &  0
  \end{pmatrix},
  \label{eq:Gamma_GG-2}
  \\
  \hat{\Gamma}^{J=0}_{Z  \gamma}
  &=
  \frac{2}{3}  \frac{\pi  \alpha_2^2}{m_V^2}
  \begin{pmatrix}
    2  c_W^2  s_W^2  &  0
    \\
    0  &  0
  \end{pmatrix},
  \label{eq:Gamma_ZG-0}
  \\
  \hat{\Gamma}^{J=2}_{Z  \gamma}
  &=
  \frac{32}{45}  \frac{\pi  \alpha_2^2}{m_V^2}
  \begin{pmatrix}
    2  c_W^2  s_W^2  &  0
    \\
    0  &  0
  \end{pmatrix},
  \label{eq:Gamma_ZG-2}
  \\
  \hat{\Gamma}^{J=0}_{Z'  \gamma}
  &=
  \frac{1}{27}  \frac{\alpha_2  g_{Z'}^2}{m_V^2}  ( 1  -  r_{Z'} )  ( 3  -  2  r_{Z'} )^2
  \begin{pmatrix}
    s_W^2  &  0
    \\
    0  &  0
  \end{pmatrix},
  \label{eq:Gamma_Zp-0}
  \\
  \hat{\Gamma}^{J=2}_{Z'  \gamma}
  &=
  \frac{8}{135}  \frac{\alpha_2  g_{Z'}^2}{m_V^2}  ( 1  -  r_{Z'} )  ( 6  +  3  r_{Z'}  +  r_{Z'}^2 )
  \begin{pmatrix}
    s_W^2  &  0
    \\
    0  &  0
  \end{pmatrix},
  \label{eq:Gamma_ZpG-2}
\end{align}
where $r_{Z'}  \equiv  \frac{m_{Z'}^2}{ 4  m_V^2 }$, and $g_{Z'}$ is defined in Eq.~\eqref{eq:g_Zp}.
We show the derivation of $\hat{\Gamma}^J$ in Appendix~\ref{sec:potential_Im}.

\subsection{Annihilation cross section}

We derive the $s$-wave spin-averaged velocity-weighted cross section for $V^0  V^0  \to  X  X'$ annihilation process through the optical theorem. 
\begin{align}
  \braket{\sigma  v_{\rm  rel}}_{XX'}
  &=  
         C  \sum_{\alpha,  \beta}  \sum_{J, J_z}  
         \left( \Gamma^{J}_{XX'} \right)_{\alpha  \beta}
         d_{2  \alpha}  ( E )  
         ~d_{2  \beta}^*  ( E ),
  \label{eq:xsec_averaged}
\end{align}
where $C=2$ for the initial states composed of the identical particles.
We introduce $E  \simeq  \frac{m_V  v^2_{\rm  rel}}{4}$ as the NR kinetic energy of the $V$-particles,
and $d_{\alpha  \beta}  (E)~(\alpha, \beta  =  1, 2)$ as the Sommerfeld enhancement factor. 
We numerically obtain $d_{\alpha  \beta}  (E)$ by solving the Schr\"{o}dinger equation~\cite{hep-ph/0412403}.

Before we show the numerical results, we remark the distinctive features of our spin-$1$ DM compared with the Wino DM, which is SU(2)$_L$ triplet spin-$\frac{1}{2}$ DM.
We focus on the leading-order results in both DM systems to show the comparison.\footnote{For the Wino DM, the Sudakov log corrections~\cite{1307.4082,1409.8294,1612.04814,1805.07367,1903.08702} are precisely evaluated.
In our spin-$1$ DM system, those potentially large corrections have not been evaluated yet, and thus our predictions have the uncertainty of ${\cal  O}  (1)$ factors.
The one-loop correction for the real part of the potential in the Wino DM system is studied in Ref.~\cite{1909.04584,2009.00640}.}
The Sommerfeld enhancement factor is approximately the same in both systems since we have the same $\hat{V}$ and the mass splitting at the leading-order.
This is because the leading-order interactions with $W$ and $Z$ bosons  in the NR limit are independent of the DM spin.
We have contributions other than $W$ and $Z$ interactions
but they are sub-leading as mentioned above.
The SM Higgs contributions for $\hat{V}$, which is zero for pure Wino DM at the leading-order, are suppressed by the small mixing angle $\phi_h$. 
The contributions from $W'$, $Z'$, and $h'$ are suppressed by their masses.\footnote{Even if we consider a relatively light $h'$ and take $m_{h'}  \sim  1$~TeV, 
the contribution to the potential is suppressed by $\exp  ( -m_{h'}/m_{Z} )  \sim  \exp  (-10)$. 
If we take a much lighter value of $m_{h'}$, 
we have to study the constraint on the $h'$ from the low energy experiments, 
which is beyond the scope of this paper.}
Consequently, we have the same resonance structure as that in the Wino system.
On the other hand, the annihilation cross section into $\gamma \gamma$ and $Z  \gamma$ are larger than those for the Wino by $\frac{38}{9}  \simeq  4.22 \cdots$. 
This is because spin-$1$ DM forms both $J  =  0$ and $J=2$ partial wave states while spin-$\frac{1}{2}$ DM forms only $J=0$ state. 
Our spin-$1$ DM also has a new annihilation mode into $Z'  \gamma $.
If this new channel is kinematically opened, a photon is emitted with the energy depending on both $m_V$ and $m_{Z'}$. 
This channel also contributes to the monochromatic gamma-ray line signals, which we study further in Sec.~\ref{sec:line_gamma-ray}.

\section{Gamma-ray line signatures}
\label{sec:line_gamma-ray}

We study the gamma-ray line signatures from our spin-$1$  DM. 
We derive the constraints from the search for monochromatic gamma-ray lines in the Galactic Center region. 
The derived constraints highly depend on the DM density profiles, and thus we show our result by taking some types of benchmark profiles to show the uncertainty.

\subsection{Line cross section}

The latest analysis for the gamma-ray line signals is performed by the H.E.S.S. collaboration using the ten years data of the gamma-ray observation in the Galactic Center region~\cite{1805.05741}. 
We constrain the gamma-ray line signatures from the DM annihilation by using this result. 
In our model, we have three annihilation modes involving the gamma-ray line signals, $\{ \gamma  \gamma,  Z  \gamma,  Z'  \gamma \}$. 
For the $\gamma  \gamma / Z  \gamma$ modes, the photon energy is approximately equal to the DM mass, 
$E_\gamma  \simeq  m_V$, where we take the NR limit for the initial DM pair and neglect $m_Z$. 
For the $Z'  \gamma$ mode, we can not neglect $m_{Z'}$ because $m_{Z'}$ must be heavier than $m_V$, see Eq.~\eqref{eq:unitarity_g0}.
The photon energy in the $Z'  \gamma$ annihilation mode depends on both $m_V$ and $m_{Z'}$.
\begin{align}
  E_{\gamma}  
  &\simeq
  m_V  \left( 1  -  \frac{m_{Z'}^2}{4  m_V^2} \right)
  \equiv  
  m_V  -  \Delta  E_\gamma, 
  \label{eq:E_gamma}
\end{align}
where $\Delta  E_\gamma  \equiv \frac{m_{Z'}^2}{4  m_V}$. 
If the $Z'  \gamma$ mode is kinematically allowed, namely for $m_{Z'}  \lesssim 2  m_V$, we have a chance to observe the double-peak gamma-ray spectrum at $E_\gamma  \simeq  m_V  -  \Delta  E_\gamma$ and $E_\gamma  \simeq  m_V$. 
To distinguish between these two peaks, $\frac{\Delta  E_\gamma}{m_V}$ should be larger than the instrumental energy resolution. 
In the H.E.S.S. experiment, the energy resolution is about $10$~$\%$ for $m_{\rm  DM}  \gtrsim  300~\text{GeV}$. 
Our interesting region is 
\begin{align}
  1.02  \lesssim  \frac{m_{Z'}}{m_V}  <  2, 
  \label{eq:mass_ratio}
\end{align}
to discuss the signal discrimination. 
The lower and upper values come from the $g_0$ perturbative unitarity and the kinematical suppression of the $Z'  \gamma$ annihilation mode, respectively. 
If we focus on this region, the condition $\frac{\Delta  E_{\gamma}}{m_V}  \gtrsim  0.1$ is always satisfied. 
Therefore, we can discriminate the gamma-ray peak originated from the $\gamma  \gamma / Z  \gamma$ modes and the peak from the $Z'  \gamma$ mode.
This double-peak gamma-ray spectrum is an outstanding feature of our DM model, and 
we can read out the values of $m_V$ and $m_{Z'}$ from this double-peak  spectrum.\footnote{Similar signals are predicted in the context of the extra-dimensional models. See  Ref.~\cite{0904.1442} for the discussion in the model with six-dimensions.}

We define the \textit{``line cross section''}, 
which contributes to the gamma-ray line signal. 
We introduce two types of line cross sections that predict the different final photon energy. 
\begin{align}
  \Braket{\sigma  v_{\rm  rel}}^{\rm  line}_{\gamma \gamma, Z  \gamma}
  &=  \Braket{\sigma  v_{\rm  rel}}_{\gamma  \gamma}  +  \frac{1}{2}  \Braket{\sigma  v_{\rm  rel}}_{Z  \gamma}
  &&
  \text{Energy peak: $E_\gamma  =  m_V$},
  \\
  \Braket{\sigma  v_{\rm  rel}}^{\rm  line}_{Z'  \gamma}
  &=  \frac{1}{2}  \Braket{\sigma  v_{\rm  rel}}_{Z'  \gamma}
  &&
  \text{Energy peak: $E_\gamma  =  m_V  \left( 1  -  \frac{m_{Z'}^2}{4  m_V^2} \right)$}. 
\end{align}
We derive the excluded region by using the experimental bound shown in Fig.~6 of Ref.~\cite{1805.05741}.
In their analysis, all the final state particles are assumed to be massless, and thus we can directly compare these constraints with $\Braket{\sigma  v_{\rm  rel}}^{\rm  line}_{\gamma \gamma, Z  \gamma}$. 
We can also derive the constraint on $\Braket{\sigma  v_{\rm  rel}}^{\rm  line}_{Z'  \gamma}$ by noting that the horizontal axis of Fig.~6 in Ref.~\cite{1805.05741} corresponds to $E_{\gamma}$. 
These constraints are derived by assuming the three cuspy DM density profiles as defined below.
\begin{itemize}
    \item  Einasto profile~\cite{0904.1442}/Einasto2 profile~\cite{1012.4515}
    \begin{align}
      \rho_{\rm  Einasto} (r)  \equiv  \rho_s  \exp  \left[ -  \frac{2}{\alpha_s}  \left( \left( \frac{r}{r_s} \right)^{\alpha_s}  -  1 \right) \right]. 
      \label{eq:Einasto}
    \end{align}
    
    \item  Navarro-Frenk-White (NFW) profile~\cite{Navarro:1996gj}
    \begin{align}
      \rho_{\rm  NFW} (r)  \equiv  \rho_s  \left( \frac{r}{r_s}  \left( 1  +  \frac{r}{r_s} \right)^2 \right)^{-1}.
    \end{align}
\end{itemize}
In Table.~\ref{tab:DM_profile}, we summarize the parameters for the cuspy DM density profiles used in Ref.~\cite{1805.05741}. 
Another choice is the cored DM density profile. 
The core radius, $r_c$, depends on the model of baryonic physics, and cores extending to $r_c \sim 5$~kpc can potentially be obtained~\cite{1405.4318}. 
The cored Einasto profile is defined as shown below. 
\begin{align}
  \rho  (r) =   
  \left\{
  \begin{array}{cc}
    \rho_{\rm  Einasto}(r)    &~~\text{for} ~~r  >  r_c,
    \\
    \rho_{\rm  Einasto}(r_c)   &~~\text{for} ~~r  <  r_c, 
  \end{array}
  \right.
\end{align}
where $\rho_{\rm  Einasto} (r)$ is defined in Eq.~\eqref{eq:Einasto}. 
For the cored profile, $\rho_s$ is chosen to realize the value of the local DM density.
We show the current excluded region with the cusped DM density profiles assumed in the analysis of the H.E.S.S. collaboration~\cite{1805.05741}.
The sensitivity of the H.E.S.S. experiment is studied for the cored profiles in Ref.~\cite{1808.04388}, which is focusing on the pure Wino DM search.\footnote{See the left panel of Fig.~7 in Ref.~\cite{1808.04388}.}
From this study, the upper bound on the line cross section will be weakened by a factor of ${\cal  O} (10-100)$
if we use the cored DM density profile. 
%
\begin{table}[tb]
  \centering
  \caption{The cuspy DM density profiles used in Ref.~\cite{1805.05741}.}
  \label{tab:DM_profile}
  
  \begin{tabular}{c | c  c  c}\hline
      Profiles    & Einasto~\cite{0904.1442}           &  NFW~\cite{Navarro:1996gj} & Einasto2~\cite{1012.4515} \\ \hline
      $\rho_{\rm  s}$  [GeV  cm$^{-3}$]  &  $0.079$  &  $0.307$  &  $0.033$
      \\
      $r_{\rm  s}$ [kpc]  &  $20.0$  &  $21.0$  &  $28.4$
      \\
      $\alpha_{\rm  s}$  &  $0.17$  &  --  &  $0.17$
      \\ 
      \hline
  \end{tabular}
\end{table}

\subsection{Constraint from gamma-ray line signatures}

\begin{figure}[t]
  \centering
  \includegraphics[width=0.8\hsize]{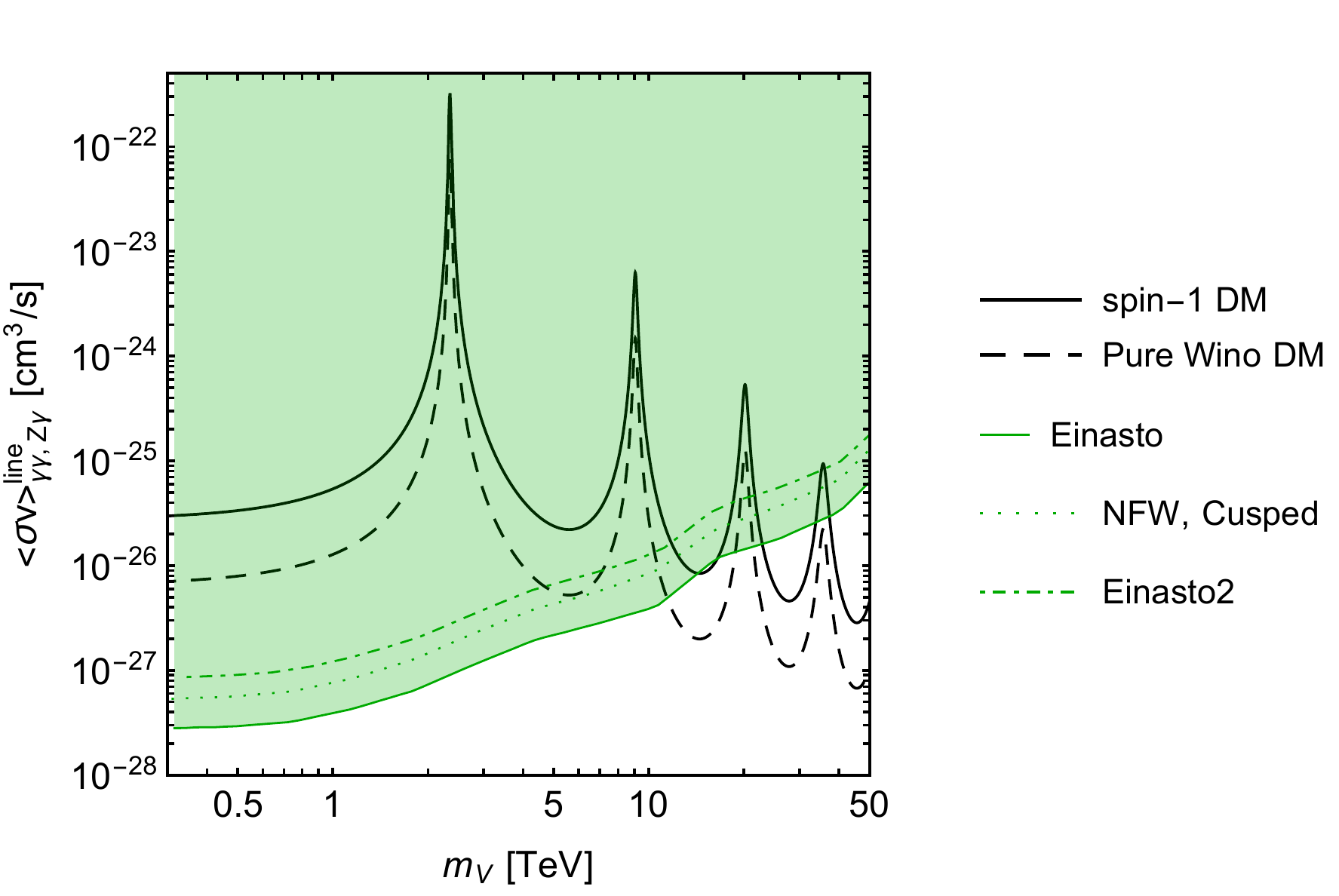}
  \caption{The comparison of the line cross sections between our spin-$1$ DM and the Wino DM. 
  The solid (dashed) black curve shows $\Braket{\sigma  v_{\rm  rel}}^{\rm  line}_{\gamma \gamma, Z  \gamma}$ for our model (pure Wino). 
  The green shaded region shows the constraint from the H.E.S.S. observation in the Galactic Center region~\cite{1805.05741} for the Einasto profile~\cite{0904.1442}. 
  The green dotted curve and dashed curve show the upper limit for the NFW profile~\cite{Navarro:1996gj} and Einasto2 profile~\cite{1012.4515}, respectively. 
  The upper bound is expected to be weakened by a factor of ${\cal  O} (10-100)$ for the cored DM density profile.
  }
  \label{fig:IDandVVGG}
\end{figure}
\begin{figure}[tb]
  \centering
  \includegraphics[width=16cm]{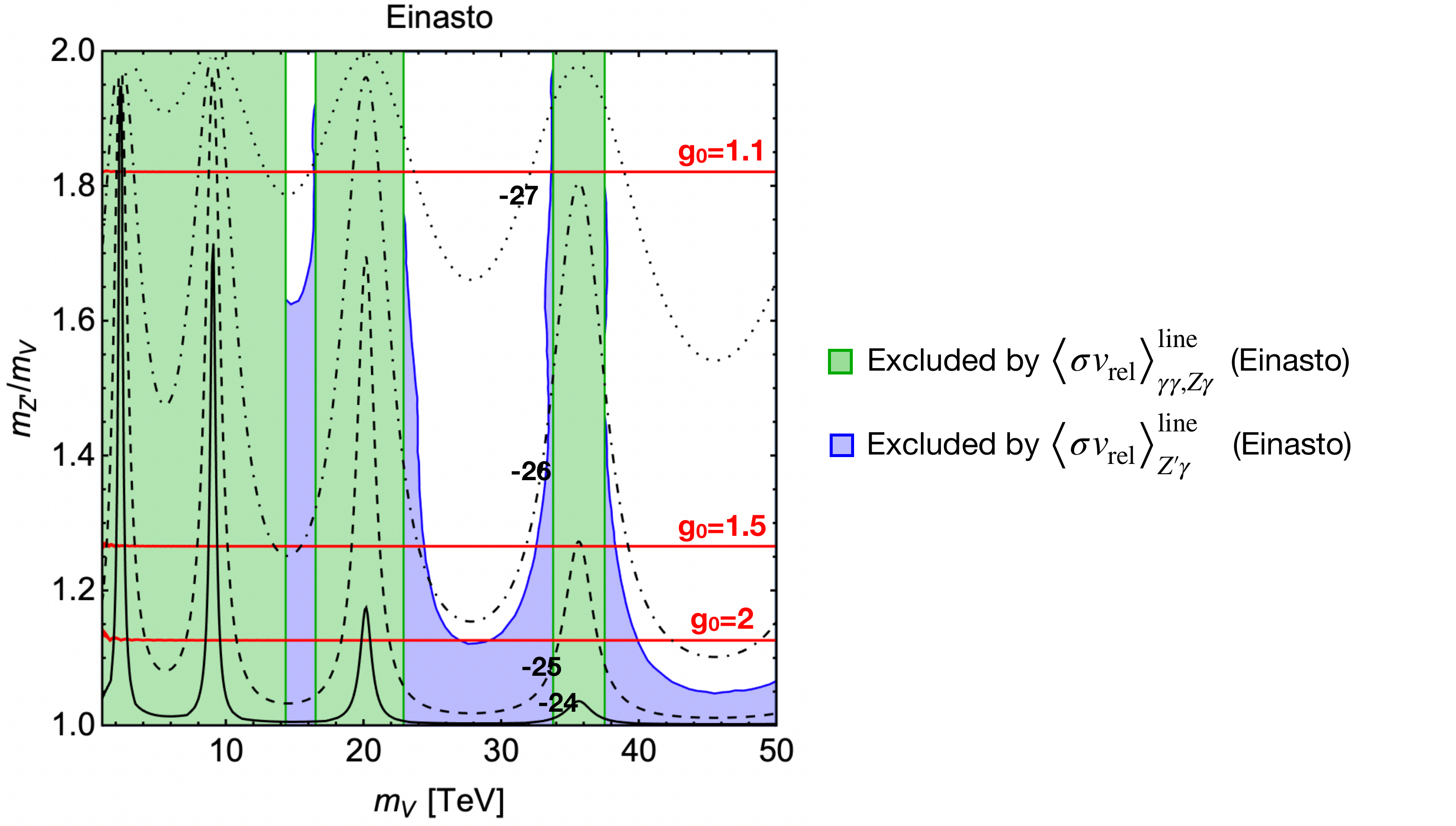}
  \caption{The current bound on the line cross section by the H.E.S.S. observation for the Einasto profile~\cite{0904.1442}. 
  The green regions are excluded by $\Braket{\sigma  v_{\rm  rel}}^{\rm  line}_{\gamma  \gamma,  Z  \gamma}$.
  The blue regions are excluded by $\Braket{\sigma  v_{\rm  rel}}^{\rm  line}_{Z'  \gamma}$.
  The black solid, dashed, dot-dashed, and dotted contours show the predicted value of $\Braket{\sigma  v_{\rm  rel}}^{\rm  line}_{Z'  \gamma}$ for $10^{-24}$, $10^{-25}$, $10^{-26}$,  and $10^{-27}$ in the unit of cm$^3$/s, respectively.  
  The red solid lines show the $g_0$ contours. 
  }
  \label{fig:excluded_region-line_Einasto}
\end{figure}

%
We compare the predicted line cross section between our spin-1 DM and the Wino system. 
In Fig.~\ref{fig:IDandVVGG}, we show the predicted value of $\Braket{\sigma  v_{\rm  rel}}^{\rm  line}_{\gamma \gamma, Z  \gamma}$ and experimental upper bounds. 
The solid (dashed) black curve shows the line cross section including Sommerfeld enhancement for our spin-$1$ DM (the pure Wino DM).
The green region with solid boundary is the excluded region by the H.E.S.S. observation~\cite{1805.05741} for the Einasto profile~\cite{0904.1442}. 
We also show green dashed and green dotted curves as the upper limits on the cross section for the cusped NFW profile~\cite{Navarro:1996gj} and the Einasto2 profile~\cite{1012.4515}, respectively.
Since our spin-$1$ DM and the Wino DM have the SU(2)$_L$ triplet features, the Sommerfeld resonance structures are almost the same. 
The line cross section for the spin-$1$ DM is larger than that for the Wino DM by $\frac{38}{9}$, and thus 
we obtain more severe constraints on the spin-$1$ DM. 
We find the following excluded regions for our spin-$1$ DM depending on the DM density profiles.\footnote{If we take $m_{h'}  \simeq  m_{V}$, 
the parameter region in Fig.~\ref{fig:excluded_region-line_Einasto} may be constrained by the perturbative unitarity bounds on the scalar couplings, which is studied in our previous collaboration~\cite{2004.00884}.
We can evade these unitarity bounds by taking sufficiently small values of $\phi_h$ and $m_{h'}$. }
\begin{itemize}
\item  Einasto profile
\begin{align}
  &
  300~\text{GeV}  \lesssim  m_V  \lesssim  14.4~\text{TeV}~~\text{(Excluded region 1)},
  \label{eq:massless_Einasto_1}
  \\
  &
  16.5~\text{TeV}  \lesssim  m_V  \lesssim  22.9~\text{TeV}~~\text{(Excluded region 2)},
  \label{eq:massless_Einasto_2}
  \\
  &
  33.8~\text{TeV}  \lesssim  m_V  \lesssim  37.5~\text{TeV}~~\text{(Excluded region 3)}.
  \label{eq:massless_Einasto_3}
\end{align}

\item  Cusped NFW profile
\begin{align}
  &
  300~\text{GeV}  \lesssim  m_V  \lesssim  12.5~\text{TeV}~~\text{(Excluded region 1)},
  \label{eq:massless_NFW_1}
  \\
  &
  17.9~\text{TeV}  \lesssim  m_V  \lesssim  22.2~\text{TeV}~~\text{(Excluded region 2)},
  \label{eq:massless_NFW_2}
  \\
  &
  34.8~\text{TeV}  \lesssim  m_V  \lesssim  36.7~\text{TeV}~~\text{(Excluded region 3)}.
  \label{eq:massless_NFW_3}
\end{align}

\item  Einasto2 profile
\begin{align}
  &
  300~\text{GeV}  \lesssim  m_V  \lesssim  11.7~\text{TeV}~~\text{(Excluded region 1)},
  \label{eq:massless_Einasto2_1}
  \\
  &
  18.6~\text{TeV}  \lesssim  m_V  \lesssim  21.6~\text{TeV}~~\text{(Excluded region 2)},
  \label{eq:massless_Einasto2_2}
  \\
  &
  35.0~\text{TeV}  \lesssim  m_V  \lesssim  36.1~\text{TeV}~~\text{(Excluded region 3)}.
  \label{eq:massless_Einasto2_3}
\end{align}
\end{itemize}
The lower value of $300$~GeV comes from the limitation for the energy resolution in the H.E.S.S experiment.

%
\begin{figure}[tb]
  \centering
  \includegraphics[width=17cm]{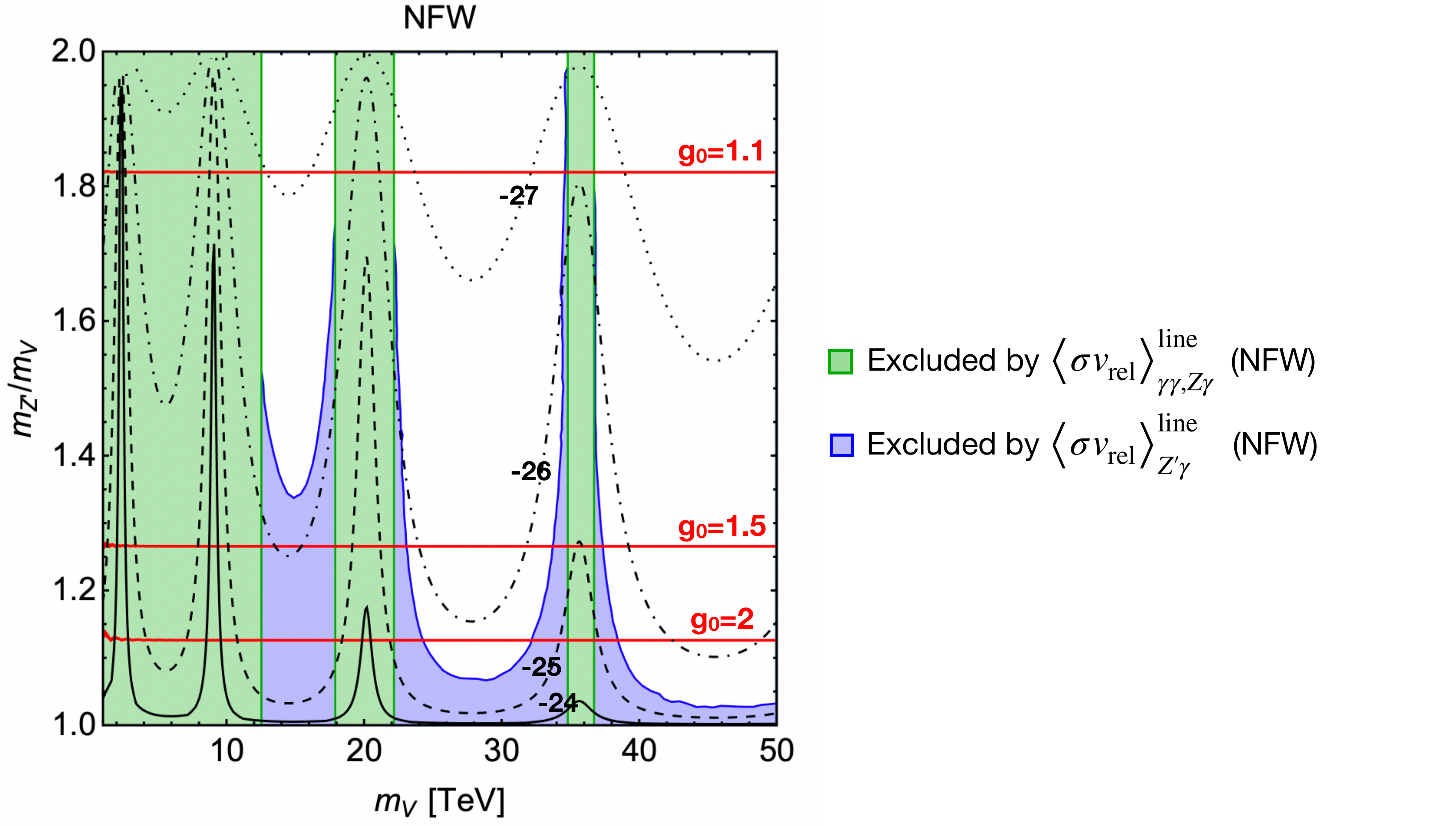}
  
  \vspace{0.5cm}
  \includegraphics[width=17cm]{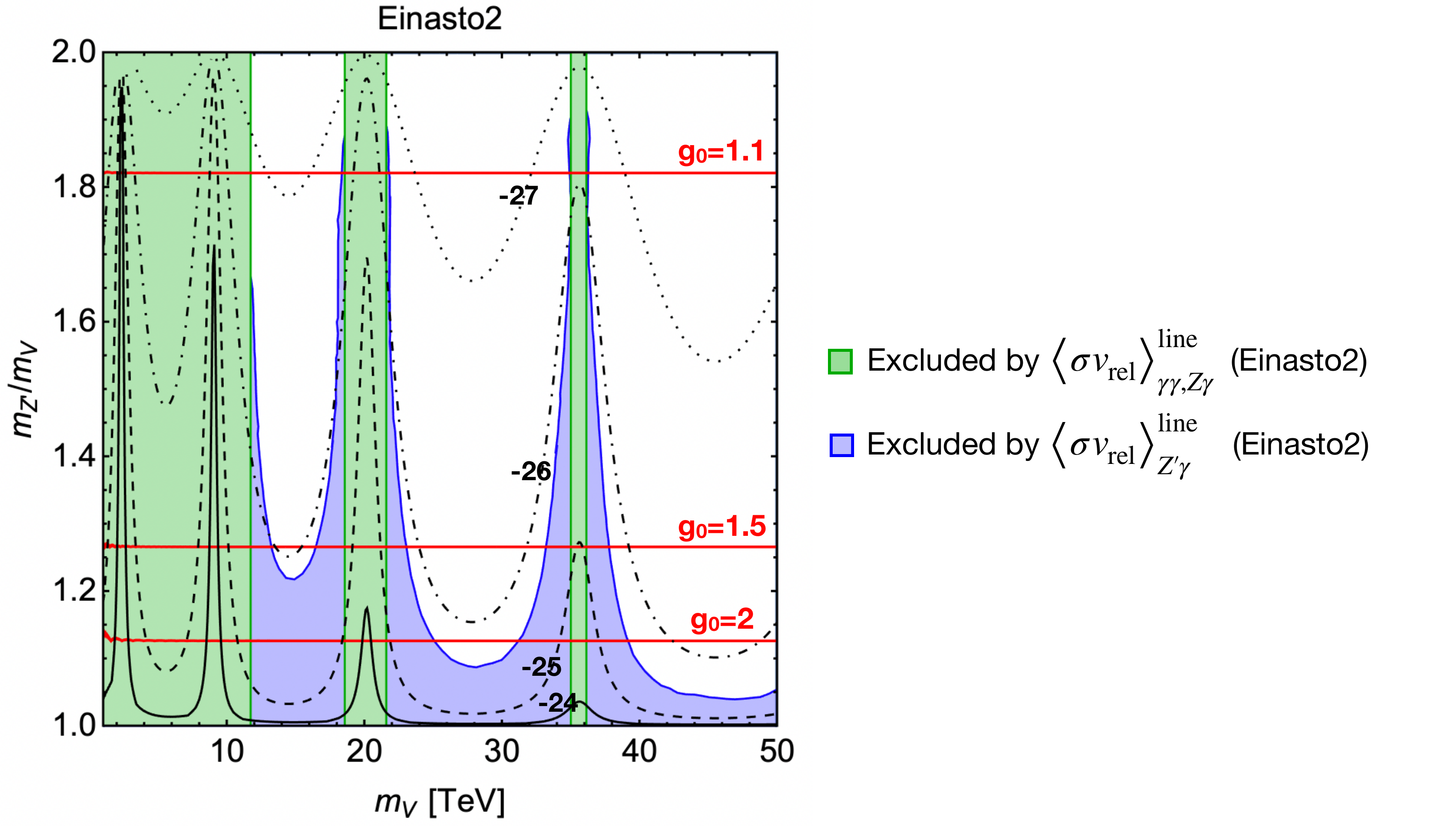}

  \caption{The constraints on the line cross section for the different DM density profiles. 
  The upper and lower panels show the constraints for the NFW profile~\cite{Navarro:1996gj} and the Einasto2 profile~\cite{1012.4515}, respectively. 
  The explanations for each plot are given in the caption of Fig.~\ref{fig:excluded_region-line_Einasto}.
  }
  \label{fig:excluded_region-line_NFW_Einasto2}
\end{figure}
%
In Fig.~\ref{fig:excluded_region-line_Einasto}, we show the current bound on the line cross section by the H.E.S.S. observation for the Einasto DM density profiles~\cite{0904.1442} 
focusing on the region in Eq.~\eqref{eq:mass_ratio}.
The green regions are excluded by $\Braket{\sigma  v_{\rm  rel}}^{\rm  line}_{\gamma \gamma, Z  \gamma}$ shown in Eqs.~\eqref{eq:massless_Einasto_1}-\eqref{eq:massless_Einasto_3}. 
These green regions are extended for $\frac{m_{Z'}}{m_V}  \geq  2$ where the annihilation into $Z'  \gamma$ is forbidden kinematically.
The blue region is excluded by $\Braket{\sigma  v_{\rm  rel}}^{\rm  line}_{Z'  \gamma}$.
The black solid, dashed, dot-dashed, and dotted contours show the predicted value of $\Braket{\sigma  v_{\rm  rel}}^{\rm  line}_{Z'  \gamma}$ for $10^{-24}$, $10^{-25}$, $10^{-26}$,  and $10^{-27}$ in the unit of cm$^3$/s, respectively.  
The red solid lines show the $g_0$ contours. 
In Fig.~\ref{fig:excluded_region-line_NFW_Einasto2}, 
we show the constraints for the NFW~\cite{Navarro:1996gj}, and Einasto2 DM density profiles~\cite{1012.4515}. 
The derived constraints depend on the DM density profiles.
The blue excluded regions from $\Braket{\sigma  v_{\rm  rel}}^{\rm  line}_{Z'  \gamma}$ give stronger constraints for $m_{Z'}  \simeq  m_V$. 
This is because we have the enhancement factor in the coupling of $Z'$, $g_{Z'}$, defined in  Eq.~\eqref{eq:g_Zp}.
Note that $g_0$ gets larger in the same region, and thus we expect relatively large higher-order correction for our perturbative calculations.

\begin{figure}[tb]
  \centering
  \includegraphics[width=16cm]{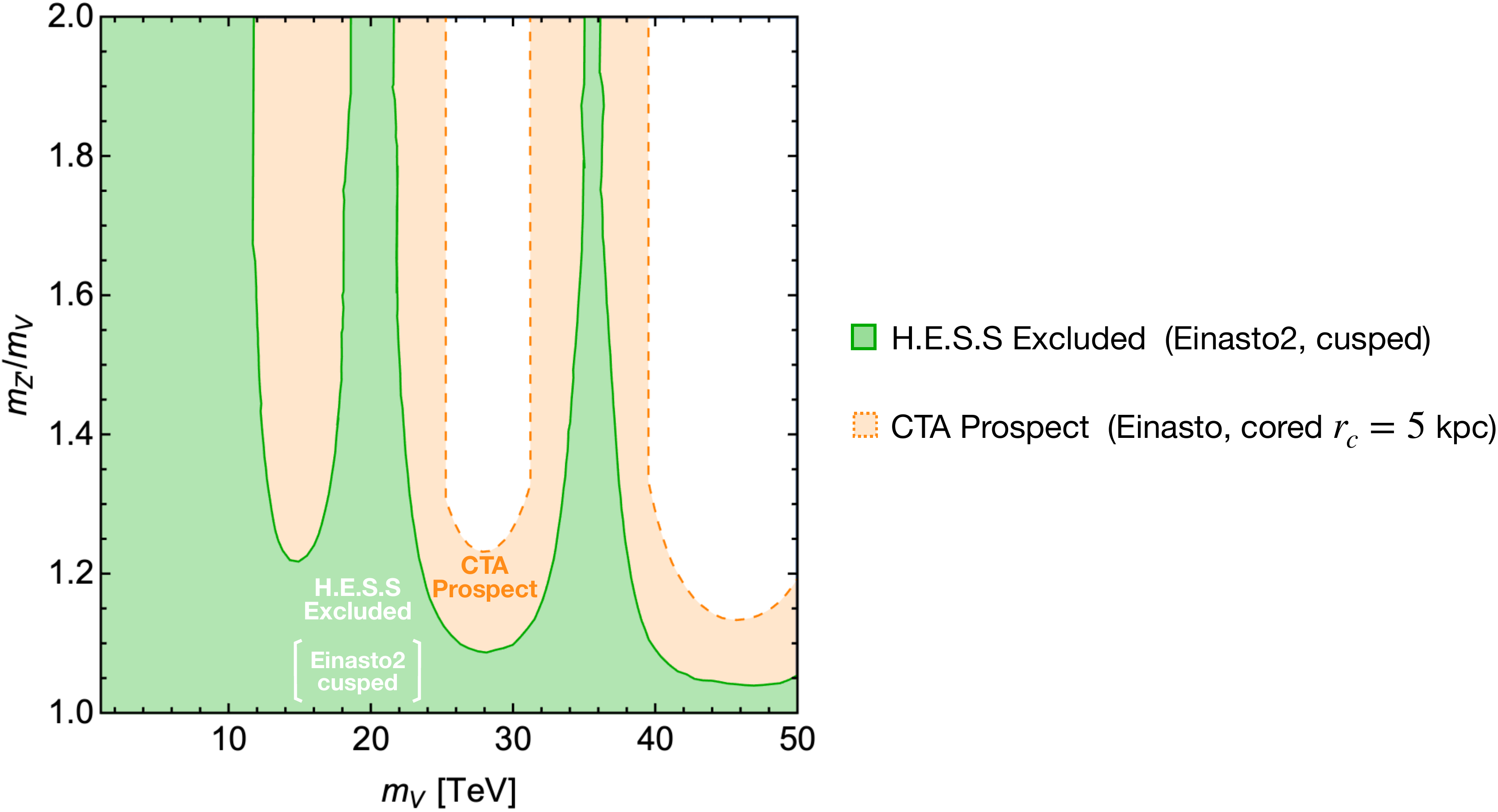}
            
  \caption{The current bound from the H.E.S.S. observation and the prospect in the CTA experiment. 
  The green region is excluded by the H.E.S.S. collaboration for the Einasto2 profile~\cite{1012.4515} shown in the lower panel of Fig.~\ref{fig:excluded_region-line_NFW_Einasto2}.
  The orange region with the dashed boundary is the prospect sensitivity in the CTA experiment for the cored Einasto profile with $r_c  =  5$~kpc, which is the most conservative one~\cite{2008.00692}. 
  If we take $r_c  \lesssim  2$~kpc, the prospect sensitivity will cover the whole region of this figure.}
  \label{fig:prospect_region-line}
\end{figure}

We have a further chance to explore the parameter region in the upcoming Cherenkov Telescope Array (CTA)~\cite{1208.5356,2007.16129}. 
The prospect sensitivity for the line gamma-ray signals is studied by Ref.~\cite{2008.00692} for the Wino and Higgsino DM.
In Fig.~\ref{fig:prospect_region-line}, we show the current bound and the future sensitivity expected in the CTA. 
The green region is excluded by the H.E.S.S. observation for the cusped Einasto2 profile shown in the lower panel of Fig.~\ref{fig:excluded_region-line_NFW_Einasto2}.
We use the prospect bound derived in Ref.~\cite{2008.00692} to show the CTA sensitivity.
The orange region with the dashed boundary shows the most conservative sensitivity assuming the core radius $r_c  =  5$~kpc, and we will obtain $m_{V}  \gtrsim  25.3$~TeV as the prospect bound. 
The whole mass range in Fig.~\ref{fig:prospect_region-line} will be covered if we take $r_c \lesssim 2$~kpc. 
%

\section{Conclusions}
\label{sec:conclusions}

In this paper, we study the gamma-ray line signatures from the electroweakly interacting non-abelian vector DM model. 
We derive the two-body effective action for the spin-$1$ DM in the NR limit. 
Since our DM has the SU(2)$_L$ triplet-like features, the Sommerfeld enhancement factor is almost the same as that for the pure Wino DM. 
The predicted annihilation cross sections into $\gamma  \gamma / Z  \gamma$ are larger by $\frac{38}{9}$ than those of the Wino. 
This is because our spin-$1$ DM pair has the additional partial wave contributions with the total spin angular momentum $J=2$. 
Therefore, the gamma-ray line signals provide stronger constraints on our spin-$1$ DM compared with the Wino DM.

We show the constraints on the line cross section for the $\gamma  \gamma$, $Z  \gamma$, and $Z'  \gamma$ modes by the H.E.S.S. observation in the Galactic Center region. 
The photon from the $\gamma  \gamma/ Z  \gamma$ modes and the $Z'  \gamma$ mode are separable through the final photon energy. 
Therefore, we may observe a double-peak in the gamma-ray energy spectrum, which provides a unique signature of our spin-$1$ DM. 
The constraints strongly depend on the DM density profiles, and we obtain $m_V  \gtrsim  14.4$~TeV ($11.7$~TeV) for the Einasto (Einasto2) profile. 
The annihilation into $Z'  \gamma$ also gives a strong constraint for $m_V  \simeq  m_{Z'}$ where couplings between DM and $Z'$ are enhanced. 
We also show the future sensitivity in the CTA experiment. 
We can probe $m_V  \gtrsim  25.3$~TeV even if we take the conservative cored DM density profile with the core radius of $5$~kpc.

We have possible extensions of our study. 
We can derive a more robust constraint on our model 
by studying the continuous gamma-ray search in observations of dwarf spheroidal galaxies by the Fermi-LAT experiment~\cite{1310.0828, 1812.06986}. 
We also expect a viable constraint from the antiproton observations as studied for the pure Wino DM~\cite{1504.05554}. 
We need to follow the decay chain including the decay of the $W'/Z'$ bosons  in our model. 
Although we expect the strongest bound comes from the gamma-ray line signatures studied in this paper, we suffer a large uncertainty in the DM density profiles. 
Therefore, it is worth evaluating the more robust constraints from the other channels. 
We also expect the Sommerfeld enhancement affects the calculations of the thermal relic abundance, which we need to derive the effective action for the electrically charged two-body states~\cite{hep-ph/0610249,2009.00640}. 
These studies are completed in our future collaborations.

\section*{Acknowledgments}

\noindent
JH thanks Seong Chan Park for useful discussions.
This work was supported by JSPS KAKENHI Grant Numbers 19H04615 and 21K03549 (TA), and JSPS Grant-in-Aid for Scientific Research KAKENHI Grant Numbers JP20J12392 (MF), JP20H01895 (JH), and JP21K03572 (JH). 
The work is also supported by JSPS Core-to-Core Program (grant number:JPJSCCA20200002).
The work of JH was supported by Grant-in-Aid for Scientific research from the Ministry of Education, Science, Sports, and Culture (MEXT), Japan (Grant Numbers 16H06492). The work of JH was also supported by World Premier International Research Center Initiative (WPI Initiative), MEXT, Japan.

\clearpage

\appendix

\section{Derivation of effective action}
\label{sec:derivation_S_eff}

We give the derivation of the effective action shown in Sec.~\ref{sec:two-body_action}.

\subsection{Real part of potential}
\label{sec:potential_Re}

\begin{figure}[htb]
  \centering
  \includegraphics[width=1\hsize]{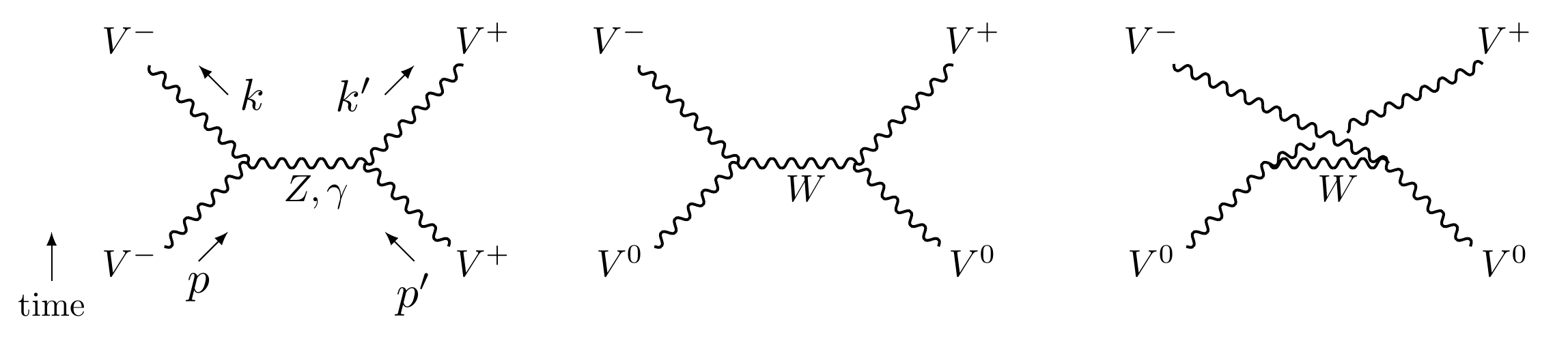}

  \caption{The relevant tree-level diagrams to derive the real part of the potential.
  The diagrams for $V^-  V^+  \to  V^0  V^0$ mode are implicit.
  }
  \label{fig:diagram-scattering}
\end{figure}
%
The real part of the potential is derived to realize the NR amplitudes of the $V$-particles. 
The leading-order contributions are induced from the diagrams shown in Fig.~\ref{fig:diagram-scattering}. 
Compared with the leading-order contributions, the Higgs exchange contributions are suppressed by small $\phi_h$. 
The $W'$ and $Z'$ exchange contributions are exponentially suppressed by $m_{W'}$ and $m_{Z'}$.
The contributions from the vector quadratic couplings are suppressed by $\frac{1}{m_V^2}$. 
In the NR limit, we obtain the following amplitude at the leading-order. 
\begin{align}
  i  {\cal  M}_{Z, \gamma}
  &\simeq
  i  4  m_V^2
  \left(
    \frac{e^2}{|\bm{p}  -  \bm{p}'|^2}
    +
    \frac{g_W^2  c_W^2}{|\bm{p}  -  \bm{p}'|^2  +  m_Z^2}
  \right)
  \epsilon_i  (p)  \epsilon_i^*  (k)  \epsilon_j  (p')  \epsilon_j^*  (k')
  ,
  \\
i  {\cal  M}_{W}
  &\simeq
  i  4  m_V^2
  \Biggl(
    \frac{g_W^2}{|\bm{p}  -  \bm{p}'|^2  +  m_W^2}
  \epsilon_i  (p)  \epsilon_i^*  (k)  \epsilon_j  (p')  \epsilon_j^*  (k')
  \nn
  \\
  &~~~~~~~~~~~~~~
    +
    \frac{g_W^2}{|\bm{p}  -  \bm{k}'|^2  +  m_W^2} 
    \epsilon_i  (p)  \epsilon_i^*  (k')  \epsilon_j  (p')  \epsilon_j^*  (k)
  \Biggr),
\end{align}
where ${\cal  M}_{Z, \gamma}$ and ${\cal  M}_{W}$ corresponds to the neutral and charged boson exchange processes, respectively.
The labels of the polarization are implicit. 
We obtain the effective action composed of the NR $V$-particle operators, which are defined in Eqs.~\eqref{eq:NR_operator-A}-\eqref{eq:NR_operator-D}, to realize the above amplitudes. 
\begin{align}
  S_{\rm  eff}
  &=
  \int  d^4  R  d^3  r
  \frac{\alpha_2  s_W^2  +  \alpha_2  c_W^2  e^{-  m_Z  r}}{r}
  \nn
  \\
  &~~~~~~~~~~~~
  \times
  \left[ 
    {\cal  B}_i^\dagger  ( R^0,  \bm{R}  +  \bm{r}/2 )~
    {\cal  B}_i  ( R^0,  \bm{R}  +  \bm{r}/2 )~
  \right]
  \left[
    {\cal  D}_j^\dagger  ( R^0,  \bm{R}  -  \bm{r}/2 )~
    {\cal  D}_j  ( R^0,  \bm{R}  -  \bm{r}/2 )
  \right]
  \nn
  \\
  &~~~~
  +
  \Biggl\{
  \int  d^4  R  d^3  r
  \frac{\alpha_2  e^{-  m_W  r}}{r}
  \nn
  \\
  &~~~~~~~~~~~~
  \times
  \left[
    {\cal  A}_i^\dagger  ( R^0,  \bm{R}  +  \bm{r}/2 )~
    {\cal  B}_i  ( R^0,  \bm{R}  +  \bm{r}/2 )
  \right]
  \left[
    {\cal  A}_j^\dagger  ( R^0,  \bm{R}  -  \bm{r}/2 )~
    {\cal  D}_j  ( R^0,  \bm{R}  -  \bm{r}/2 )
  \right]
  \nn
  \\
  &~~~~~~~~~~~~
  +  h.c. 
  \Biggr\}. 
\end{align}
To express the action in the decomposed form into each partial wave mode, 
we reform the vector indices ($i$, $j$) by using the Fierz identity for Grassmann-even operators. 
\begin{align}
  \delta_{i  j}  \delta_{k  \ell}  =  \sum_{J, J_z}  (-1)^J  S_{i  \ell}^{J,  J_z}  S_{k  j}^{J,  J_z  *}
  ~~~~~~( i,  j,  k,  \ell  =  1,  2,  3 ), 
\end{align}
where $S_{i  j}^{J,  J_z}$ is defined in Eqs.~\eqref{eq:Sij_def-0}-\eqref{eq:Sij_def-2}. 
After this decomposition, we can express the action in terms of the two-body states defined in Eqs.~\eqref{eq:phi_C}-\eqref{eq:phi_N} and read out $\hat{V}$ as given in Eq.~\eqref{eq:V_0-2}.

\subsection{Imaginary part of potential}
\label{sec:potential_Im}

\subsubsection{Matching procedure}

To derive the imaginary part of the potential, $\hat{\Gamma}^J$, we perform the operator matching between the two-body field operators and the calculations of the one-loop amplitudes of the $V$-particles.
Since we focus on the annihilation modes into the neutral vectors, the $(1,1)$-component of $\hat{\Gamma}^J$ only has the nonzero value. 
Therefore, we can use the optical theorem to calculate the imaginary part of the one-loop amplitude. 
\begin{align}
  \left.
  {\rm  Im}  {\cal  M}
  \right|_{ V^-  V^+  \to  X  X'  \to  V^-  V^+ }^{J}
  &=
  2  m_V^2  
  \Braket{\sigma  v_{\rm  rel}}^{J}_{V^-  V^+  \to  X  X'}
  ~~~~(XX'  =  \gamma  \gamma, Z  \gamma, Z'  \gamma)
  ,
  \label{eq:optical_theorem-pm}
\end{align}
where the left-hand side denotes the imaginary part of the forward scattering amplitude for $V^-  V^+  \to  X  X'  \to  V^-  V^+$  with $J$ which is expressed by $(\hat{\Gamma}^{J}_{X  X'})_{11}$. 
In the right-hand side, $\Braket{\sigma  v_{\rm  rel}}^{J}_{V^-  V^+  \to  X  X'}$ denotes the  partial wave annihilation cross section of $V^-  V^+  \to  X  X'$ for the initial state with $J$.
In the evaluation of the annihilation cross section, we only leave the leading-order terms in the NR limit
and take the massless limit for all the SM particles in the final states while we leave the masses of $Z'$. 
We derive $\hat{\Gamma}^J$ through the above procedure
as summarized in Eqs.~\eqref{eq:Gamma_GG-0}-\eqref{eq:Gamma_ZpG-2}. 
In the succeeding section, we show how to determine $\hat{\Gamma}^{J}_{\gamma  \gamma}$ 
as a demonstration.

\subsubsection{Derivation of $~\hat{\Gamma}_{\gamma  \gamma}^{J}$}

\begin{figure}[htb]
  \centering
  \includegraphics[width=1\hsize]{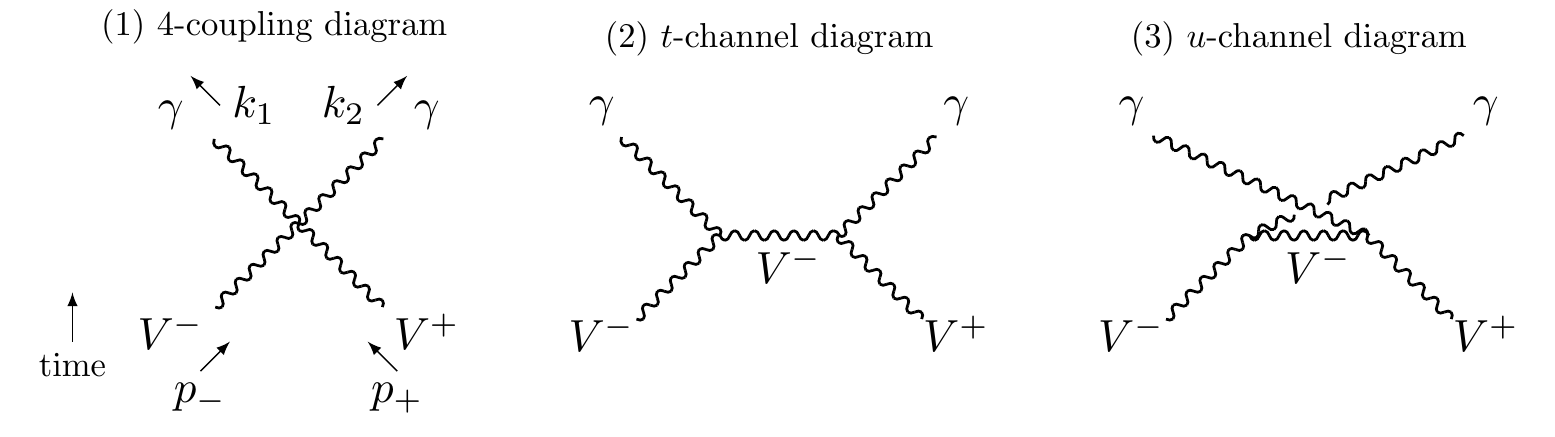}

  \caption{The tree-level diagrams which contribute to $V^- V^+ \to \gamma \gamma$ annihilation.
  }
  \label{fig:diagram-annihilation}
\end{figure}
%
We show the derivation of $\hat{\Gamma}_{\gamma  \gamma}^{J}$ as a concrete example.
We can focus on the nonzero component, $(\hat{\Gamma}_{\gamma  \gamma}^{J})_{11}$. 
First, we calculate the velocity-weighted annihilation cross section for $V^-  V^+  \to  \gamma  \gamma$.
The tree-level diagrams in the unitarity gauge are shown in Fig.~\ref{fig:diagram-annihilation}.
The amplitudes that correspond to each diagram are obtained as follows.
\begin{align}
  i \mathcal{M}_4 
  =& ie^2(g^{\mu \rho} g^{\nu \sigma} + g^{\nu \rho}g^{\mu \sigma} -2 g^{\mu \nu}g^{\rho \sigma}) 
        \epsilon_\sigma (p_-) \epsilon_\rho (p_+) \epsilon^*_\mu (k_1) \epsilon^*_\nu (k_2),
        \\
  i \mathcal{M}_t 
  =& ie[(p_-+q)^\mu g^{\alpha \sigma} + (-p_--k_1)^\alpha g^{\mu \sigma}+(k_1-q)^\sigma g^{\mu \alpha}  ]
       \frac{-i}{q^2 - m_V^2} \left( g_{\alpha \beta} - \frac{q_\alpha q_\beta}{m_V^2} \right) \cr
       &
       \times ie[ (q-p_+)^\nu g^{\beta \rho} + (-q-k_2)^\rho g^{\beta \nu} + (k_2 + p_+)^\beta g^{\rho \nu} ] 
       \epsilon_\sigma (p_-) \epsilon_\rho (p_+) \epsilon^*_\mu (k_1) \epsilon^*_\nu (k_2), 
       \\
  i \mathcal{M}_u 
  =& i e[ (p_-+q')^\nu g^{\alpha \sigma} + (-p_--k_2)^\alpha g^{\nu \sigma} +(k_2-q')^\sigma g^{\nu \alpha}  ]
        \frac{-i}{q'^2 - m_V^2} \left( g_{\alpha \beta} - \frac{q'_\alpha q'_\beta}{m_V^2} \right) \cr
        &
        \times ie[ (q'-p_+)^\mu g^{\beta \rho} + (-q'-k_1)^\rho g^{\beta \mu} + (k_1 + p_+)^\beta g^{\rho \mu} ] 
        \epsilon_\sigma (p_-) \epsilon_\rho (p_+) \epsilon^*_\mu (k_1) \epsilon^*_\nu (k_2),
\end{align}
where $q \equiv p_- - k_1$ and $q' \equiv p_- - k_2$.

In the NR limit for the initial $V$-particles, the zeroth components of the polarization vectors are sub-leading, 
and we focus on the spatial component, $\epsilon_i (p_-) \epsilon_j (p_+)$.
We decompose the amplitude into each partial wave contribution by replacing the initial state polarization vectors with $S^{J,  J_z}_{ij}$ defined in Eqs.~\eqref{eq:Sij_def-0}-\eqref{eq:Sij_def-2}.
\begin{align}
     \epsilon_i (p_-) \epsilon_j (p_+)  \to  S^{J,J_z}_{i  j}. 
\end{align}
Note that $S^{J,J_z}_{ij}$ is symmetric  matrices for $J=0,2$ and anti-symmetric for $J=1$. 
Thanks to these properties, the amplitudes in the center-of-mass frame are obtained in the decomposed form,~${\cal  M}^{J,J_z}$.
\begin{align}
  \mathcal{M}^{0,0} 
  =& 2e^2 
        \left[ 
          4 \epsilon_i^*  (k_1) S^{0,0}_{ij} \epsilon_j^*  (k_2) 
          - \epsilon_i^*  (k_1) \epsilon_i^*  (k_2) \left( S^{0,0}_{ii}
          - \frac2{m_V^2} k_j  S^{0,0}_{jk} k_k \right) 
        \right], 
        \\
  \mathcal{M}^{1,J_z} 
  =& 0,
  \\
  \mathcal{M}^{2,J_z} 
  =& 2e^2 
        \left[ 
          4 \epsilon_i^*  (k_1) S^{2,J_z}_{ij} \epsilon_j^*  (k_2) 
          +\epsilon_i^*  (k_1)\epsilon_i^*  (k_2)  \frac2{m_V^2} k_j S^{2,J_z}_{jk} k_k 
        \right],
\end{align}
where $|J_z|  \leq  J$ and $k_i$ denotes the spatial components of ${k_1}_\mu$. 
We use $S^{2,J_z}_{ii}  =  0$ to obtain the above result. 
The polarization vectors for the photons are given by
\begin{align}
  \epsilon^\pm_i (k_1) 
    &=\frac1{\sqrt2} \begin{pmatrix} \mp \cos\theta\\-i\\\pm \sin\theta \end{pmatrix},
    &
  \epsilon^\pm_i (k_2) 
    &= \frac1{\sqrt2} \begin{pmatrix} \pm \cos\theta\\-i\\\mp \sin\theta \end{pmatrix},
\end{align}
where $\theta$ is the scattering angle for the photon with its momentum $k_1$.

After taking the sum over the final state spins, the squared amplitudes can be expressed as follow. 
\begin{align}
  \sum_{\rm  final} | \mathcal{M}^{0,0} |^2 &= 24 e^4, 
  \\
  \sum_{\rm  final} | \mathcal{M}^{1,J_z} |^2 &= 0
  ~~~~~~~~~~~~~~~~~~~~~~~~~~~~~~~~~~~~~~~~~
  (J_z  =  0,  \pm  1),
  \\
  \sum_{\rm  final} | \mathcal{M}^{2,J_z} |^2 &= 
    \left\{
      \begin{array}{ll}
        48e^4 \sin^4\theta & \hspace{20pt} (J_z=0), \\
        16e^4 \sin^2\theta(3 +\cos2\theta)&\hspace{20pt} (J_z=\pm1), \\
        8e^4 (1+6\cos^2\theta+\cos^4\theta)&\hspace{20pt} (J_z=\pm2). 
      \end{array} 
    \right.
\end{align}
We obtain the cross section for each partial wave mode.
\begin{align}
    \Braket{\sigma  v_{\rm rel}}^{J=0}_{V^-  V^+  \to  \gamma  \gamma} &= 6 \frac{\pi \alpha_2^2}{m_V^2}  s_W^4,
    \\
    \Braket{\sigma  v_{\rm rel}}^{J=1}_{V^-  V^+  \to  \gamma  \gamma} &= 0,
    \\
    \Braket{\sigma  v_{\rm rel}}^{J=2}_{V^-  V^+  \to  \gamma  \gamma} &= \frac{32}{5} \frac{\pi \alpha_2^2}{m_V^2}  s_W^4.
\end{align}
The spin-averaged total cross section is obtained by adding up all the partial wave cross sections.
\begin{align}
    \overline{\Braket{\sigma  v_{\rm rel}}}^{\rm  tot}_{V^-  V^+  \to  \gamma  \gamma}
    &= \frac{38}9 \frac{\pi \alpha^2_2}{m_V^2}  s_W^4.
\end{align}
This cross section is larger than that of the Wino DM by a factor of $\frac{38}{9}$. 
For the $Z  \gamma$ mode, we take the massless limit of the $Z$ boson, and the calculation procedure is the same.
For the $Z'  \gamma$ mode, we do not neglect $m_{Z'}$, and the longitudinal mode also contributes to the final result.

Next, we calculate the imaginary part of the forward scattering amplitude  using the two-body effective action shown in Eq.~\eqref{eq:SII}.
Namely, we use the following coupling to calculate the $\gamma  \gamma$ contribution to ${\rm  Im} {\cal  M}$.
\begin{align}
  S_{\rm  eff}
  &\supset
  \sum_{J, J_z}
  i  \frac{9}{2}  (\hat{\Gamma}_{\gamma  \gamma}^{J})_{11}  
  \int  d^4  R
  ~ 
  \phi_C^{J, J_z \dagger}  ( R,  \bm{0} )  
  ~
  \phi_C^{J, J_z}  ( R,  \bm{0} ).
\end{align} 
Using this coupling, we obtain
\begin{align}
  \left.
  {\rm  Im}  {\cal  M}
  \right|_{ V^-  V^+  \to  \gamma  \gamma  \to  V^-  V^+ }^{J}
  &=
  2  m_V^2  \cdot  9  (\hat{\Gamma}^{J}_{\gamma  \gamma})_{11}.
\end{align} 
Comparing with \eqref{eq:optical_theorem-pm}, we can determine $(\hat{\Gamma}^{J}_{\gamma  \gamma})_{11}$ as shown below. 
\begin{align}
  (\hat{\Gamma}^{J}_{\gamma  \gamma})_{11}  
  &=  \frac{1}{9} \Braket{\sigma  v_{\rm  rel}}_{V^-  V^+  \to  \gamma  \gamma}^{J}
  =
      \left\{
    \begin{array}{ll}
        \frac{2}{3} \frac{\pi \alpha_2^2}{m_V^2}  s_W^4, 
        \\
        0, 
        \\
        \frac{32}{45} \frac{\pi \alpha_2^2}{m_V^2}  s_W^4. 
        \\
        \end{array} \right. 
\end{align}
This result is summarized in Eqs.~\eqref{eq:Gamma_GG-0}-\eqref{eq:Gamma_GG-2}. 
We can also determine $\hat{\Gamma}^{J}_{Z  \gamma}$ and $\hat{\Gamma}^{J}_{Z'  \gamma}$ in the same procedure.


\end{document}